\documentclass[10pt]{article}
\usepackage[margin=1in]{geometry}

\newcommand\BibTeX{{\rmfamily B\kern-.05em \textsc{i\kern-.025em b}\kern-.08em
T\kern-.1667em\lower.7ex\hbox{E}\kern-.125emX}}

\usepackage{bm}
\usepackage{natbib}
\usepackage{setspace}
\usepackage[hidelinks,colorlinks=true,linkcolor=blue,citecolor=blue]{hyperref}
\usepackage{amsthm}
\usepackage{xcolor}
\usepackage{enumitem}
\usepackage{bm}
\usepackage{amsmath}
\usepackage{amssymb}
\usepackage{graphicx}
\usepackage{amsfonts}
\usepackage{mathtools}
\usepackage{algorithm}
\usepackage{setspace}
\usepackage{graphicx}
\usepackage[
singlelinecheck=false,
labelfont=bf,font={doublespacing, small}
]{caption}

\usepackage[affil-it]{authblk} 
\usepackage{etoolbox}
\usepackage{lmodern}

\usepackage{sectsty}
\allsectionsfont{\sffamily}

\DeclareMathOperator*{\argmin}{argmin}
\DeclareMathOperator{\E}{\mathbb{E}}

\newcommand{\Var}{\mathrm{Var}}
\newcommand{\tp}{\overset{p}{\to}}
\newcommand{\td}{\Rightarrow}
\newcommand{\indep}{\perp \!\!\! \perp}

\newcommand{\pa}{\overset{\mathrm{p}}{\to}}

\newcommand{\Cov}{\mathrm{Cov}}
\newcommand{\diag}[1]{\mathrm{diag}\{#1\}}

\usepackage{apptools}
\AtAppendix{\counterwithin{lemma}{section}}
\AtAppendix{\counterwithin{theorem}{section}}
\AtAppendix{\counterwithin{remark}{section}}
\AtAppendix{\counterwithin{assumption}{section}}

\begin{document}

\newtheoremstyle{sf}% style name
{2ex}% above space
{2ex}% below space
{ }% body font
{ }% indent amount
{\scshape\sffamily\bfseries}% head font
{. }% post head punctuation
{ }% post head punctuation
{ }% head spec

\theoremstyle{sf}
\newtheorem{theorem}{Theorem}
\newtheorem{lemma}{Lemma}
\newtheorem{corollary}{Corollary}
\newtheorem{definition}{Definition}
\newtheorem{remark}{Remark}
\newtheorem{procedure}{Procedure}
\newtheorem{assumption}{Assumption}

\title{Accounting for inconsistent use of covariate adjustment in group sequential trials}
\doublespacing

\author[1]{Marlena S. Bannick}
\author[2,3]{Sonya L. Heltshe}
\author[1,3]{Noah Simon}

\affil[1]{Department of Biostatistics, University of Washington, Seattle, Washington, U.S.A.}

\affil[2]{Department of Pediatrics, University of Washington, Washington, United States}

\affil[3]{Cystic Fibrosis Foundation, Therapeutics Development Network Coordinating Center, Seattle Children's Research Institute, Washington, United States}

\maketitle

\begin{abstract}
Group sequential designs in clinical trials allow for interim efficacy and futility monitoring. Adjustment for baseline covariates can increase power and precision of estimated effects. However, inconsistently applying covariate adjustment throughout the stages of a group sequential trial can result in inflation of type I error, biased point estimates, and anti-conservative confidence intervals.  We propose methods for performing correct interim monitoring, estimation, and inference in this setting that avoid these issues. We focus on two-arm trials with simple, balanced randomization and continuous outcomes. We study the performance of our boundary, estimation, and inference adjustments in simulation studies. We end with recommendations about the application of covariate adjustment in group sequential designs.
\end{abstract}

\newpage
\section{Introduction}

Randomized controlled trials (RCTs) are often viewed as the gold standard study design for determining the efficacy of health interventions. It is important that RCTs make efficient use of resources. To that end, two techniques in the design and analysis of RCTs have become popular: group sequential designs, and covariate-adjusted analyses.

The group sequential trial design allows for early termination of a trial based on strong signals of efficacy or futility. Since investigators perform multiple analyses on the same data over time, standard group sequential designs set higher thresholds to reject the null hypothesis at each interim analysis. This carefully controls the type I error across the whole duration of the trial. Covariate adjustment refers to using pre-randomization variables in an analysis of the treatment effect between groups. When performed correctly, covariate adjustment can decrease the variance of the estimated treatment effect without changing the estimand. This increases estimate precision and power to reject the null hypothesis. Both group sequential designs and covariate adjustment -- and their use simultaneously -- have been studied extensively elsewhere \citep{yangEfficiencyStudyEstimators2001, colantuoniLeveragingPrognosticBaseline2015, vanlanckerImprovingInterimDecisions2019a, wangAnalysisCovarianceRandomized2019, wangModelRobustInferenceClinical2021, vanlanckerCombiningCovariateAdjustment2022a, yeBetterPracticeCovariate2022}.

\paragraph{Hybrid design} We study the scenario where covariate adjustment is performed inconsistently within a group sequential trial design. Specifically, we are interested in the ``hybrid design'', where un-adjusted analyses are used for interim monitoring and covariate-adjusted analyses are used to obtain a final treatment estimate. The hybrid design may arise from investigator preference or practical concerns. For example, there could be an administrative delay in making covariates available for inclusion in an analysis until the end of the study. Two recent examples in cystic fibrosis (CF) research used this hybrid design. The GROW trial investigated the effect of oral glutathione on grown in children with CF \citep{bozicOralGlutathioneGrowth2020}. The TEACH trial investigated the effect of adding oral azithromycin to inhaled tobramycin on lung function in people with CF who are also chronically infected with Pseudomonas aeruginosa \citep{nicholsTestingEffectsCombining2021}.

The classical critical values in a group sequential trial (e.g., Pocock \citep{pocockGroupSequentialMethods1977} or O'Brien-Fleming \citep{obrienMultipleTestingProcedure1979} boundaries) are predicated on conducting the same analysis at each stage of the trial. Changing the analysis strategy at the end of the trial (or any other time during the course of the trial)  without proper adjustments may result in inflation of type I error, biased point estimates, and anti-conservative confidence intervals. Even with appropriate critical values, traditional point estimates and confidence intervals can lead to incorrect inference: point estimates may be biased \citep{kimPointEstimationFollowing1989}, and confidence intervals may be anti-conservative \citep{kimConfidenceIntervalsFollowing1987}. Adjusted versions of these quantities have been proposed that take into account the standard group sequential design, but not the hybrid design \citep{kimPointEstimationFollowing1989, kimConfidenceIntervalsFollowing1987, siegmundEstimationFollowingSequential1978, gernotGroupSequentialConfirmatory2016}.

We address both of these problems. We propose adjustments for critical values to ensure correct type I error in the hybrid design. We also propose techniques to obtain median-unbiased point estimates and asymptotically correct confidence intervals in the hybrid design. These techniques mirror the traditional procedures for obtaining for point estimates and confidence intervals in group sequential designs. We justify our proposed methods theoretically. We show empirically through simulation studies that they preserve type I error and produce correct estimation and inference. We focus on two-arm trials with continuous outcomes using Analysis of Variance (ANOVA) as the un-adjusted analysis and Analysis of Covariance (ANCOVA) as the covariate-adjsuted analysis.

\section{Preliminaries}\label{meth:intro}

Our contribution is presented in Section \ref{hybrid}. In this section, we briefly review the aspects of group sequential designs and covariate adjustment that are critical building blocks for our proposed methods in Section \ref{hybrid}. We emphasize that the technical results on covariate adjustment in Lemma \ref{lemma:identification-estimation} have been shown elsewhere \citep{yangEfficiencyStudyEstimators2001, colantuoniLeveragingPrognosticBaseline2015, wangAnalysisCovarianceRandomized2019}, but we re-prove them in the Supplemental Materials for the convenience of the reader. The reader familiar with group sequential designs and covariate adjustment may choose to skip ahead to Section \ref{hybrid} after reviewing the notation in our standard assumptions and definitions. We will refer to these assumptions and definitions when we propose valid methods for interim monitoring, estimation, and inference in the hybrid design.

\paragraph{Notation} Throughout, we use capital letters to refer to random variables $X$ (bold-face $\bm X$ for random vectors), and lowercase letters to refer to scalar observations $x$ and vector or matrix observations $\bm x$. We use the notation $X_n \tp c$ when $X_n$ converges in probability to $c$, and we use $X_n \td X$ when $X_n$ weakly converges to $X$. We use the notation $o_P(r_n)$ to refer to a random variable such that if $X_n = o_P(r_n)$, $r_n^{-1} X_n \tp 0$, e.g., if $X_n = o_P(n^{-\frac{1}{2}})$, then $\sqrt{n} X_n \tp 0$. Finally, we use the notation $\bm A\otimes \bm B$ to denote the Kronecker product of two matrices $\bm A$ and $\bm B$.

\subsection{Review of group sequential designs}
\begin{figure}
    \centering
    \caption{Diagram of group sequential trial with three stages. Figure (a) shows the standard group sequential trial design with the same analysis method to obtain $Z_k^*$ at each stage $k = 1, ..., 3$. Figure (b) shows the modified group sequential trial design where covariate adjustment is only performed after crossing an interim monitoring boundary, or the end of the trial is reached.}
    \includegraphics[width=0.9\textwidth]{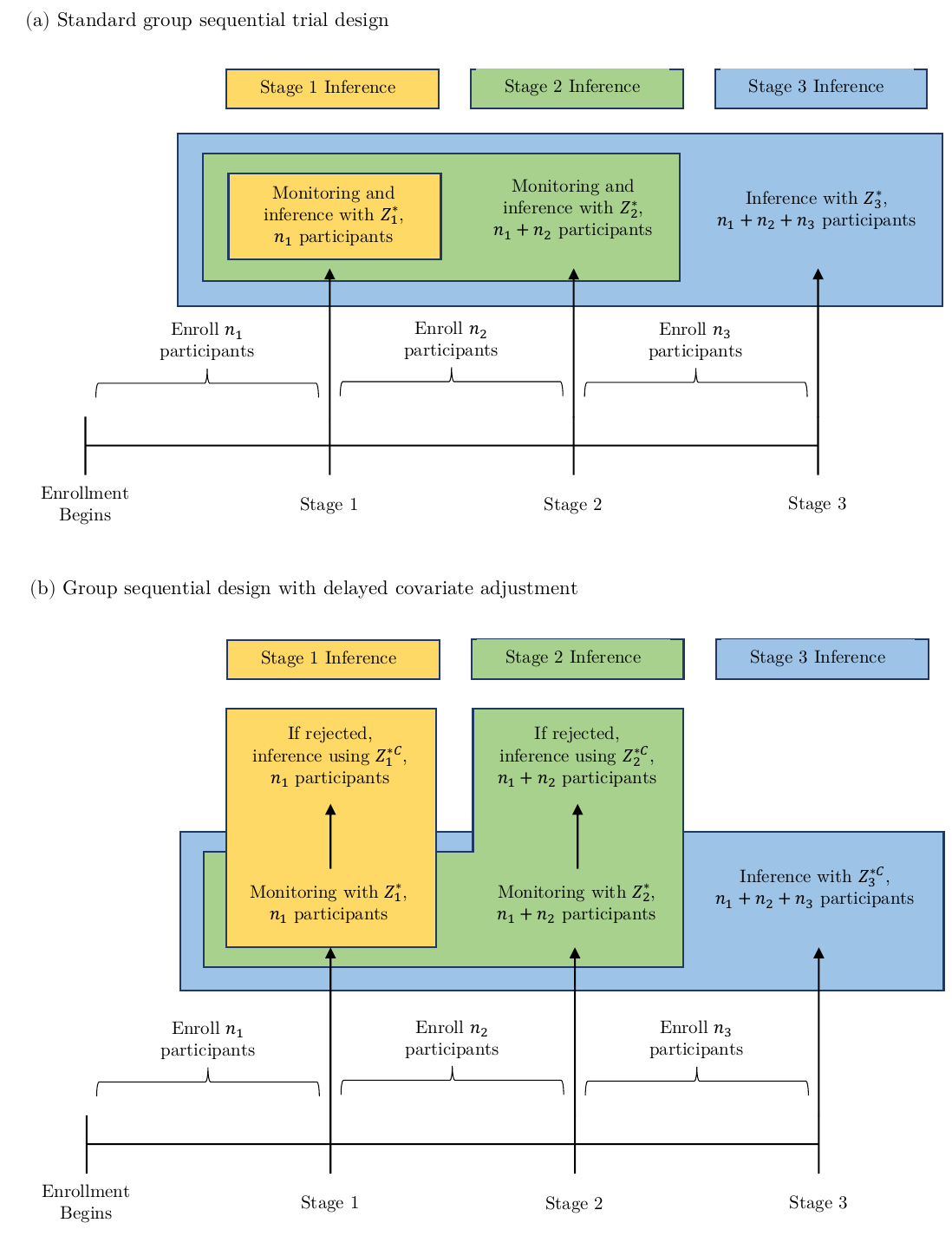}
    \label{fig:gst}
\end{figure}

In a group sequential design, individuals are enrolled sequentially across stages of a trial. Hypothesis testing is also performed sequentially, at the end of each stage of enrollment. This procedure is depicted in Figure \ref{fig:gst}(a).
Assumption \ref{assump:sequential} formally defines the asymptotic regime that we consider for a group sequential design.
\begin{assumption}[Group Sequential Asymptotic Regime]\label{assump:sequential}
   Let $K$ be a fixed, maximum number of stages in a group sequential trial, and let $n$ be the maximum sample size of the trial. Let $\{t_k; k = 1, ..., K\}$ be a set of fixed \textit{information fractions}. Consider $n \to \infty$. Then,
   \begin{itemize}
       \item $n_k = \text{floor}(n \cdot t_k)$ is the number of individuals enrolled at at stage $k$
       \item $n_k^* = \sum_{k'=1}^{k} n_k$ is the cumulative number of individuals enrolled by stage $k$
       \item $t^*_k = n_k^* / n_K$ is the cumulative information fraction.
   \end{itemize}
\end{assumption}
We perform a hypothesis test at stage $k$ using a test statistic derived using $n_k^*$ observations. The null hypothesis of no treatment effect is rejected if the test statistic exceeds stage-specific critical values $(l_k, u_k), k = 1, ..., K$. The test statistics that we consider for the hybrid design are discussed in detail in Section \ref{hybrid}.

\subsection{Review of covariate adjustment}
We first review classical results on covariate adjustment for estimating the average treatment effect in a randomized tria \citep{yangEfficiencyStudyEstimators2001, colantuoniLeveragingPrognosticBaseline2015, wangAnalysisCovarianceRandomized2019}. In Assumption \ref{assump:data} we describe the standard assumptions that we require of the data generating process. We note that some readers may be more familiar with treatment assignments labeled as $\{0, 1\}$ rather than $\{-1, 1\}$. All of the results in this and future sections still apply using $\{0, 1\}$. We choose to use $\{-1, 1\}$ as it is a common convention and eases exposition of technical proofs. Furthermore, we note that the mean-zero requirement in Assumption \ref{assump:data} (iii) can be relaxed, but it also eases exposition.
\begin{assumption}[Data Generation]\label{assump:data}
    We require the following assumptions of the data generating process. Let $(Y, A, \bm X) \sim P$, where $P$ satisfies the following:
    \begin{enumerate}[label=(\roman*)]
        \item $Y \in \mathbb{R}$ is the continuous outcome of interest
        \item $A \in \{-1, 1\}$ is the treatment assignment, and has a Rademacher distribution, i.e., $P(A = -1) = P(A = 1) = 0.5$
        \item $\bm X = (X_1, ..., X_p) \in \mathbb{R}^p$ are $p$ covariates such that $A \indep \bm X$
    \end{enumerate}
\end{assumption}
Our target of inference is the average treatment effect (ATE), defined in Definition \ref{def:ate}.
\begin{definition}[Average Treatment Effect]\label{def:ate}
We define the following as the average treatment effect:
\begin{align*}
    \mathrm{ATE}: \quad \E_P[Y|A = 1] - \E_P[Y|A=-1].
\end{align*}
\end{definition}
Consider the following population minimizer of squared error loss for predicting $Y$ with $A$, $\Delta^*$:
\begin{align}\label{dstar}
    (\Delta^{*}, \theta^{*}) = \argmin_{\Delta, \theta \in \mathbb{R}} \E_P[(Y - \theta - \Delta A)^2]
\end{align}
Now consider the case where we have explained some of the variation in the outcome $Y$ with covariates $\bm X$ using a population line of best fit:
\begin{align}\label{dstar-cov}
    (\Delta^{*}_C, \theta^{*}_C, \bm \gamma^{*}_C)^T = \argmin_{\Delta, \theta\in\mathbb{R}, \bm \gamma \in \mathbb{R}^{p}} \E_P[(Y - \theta - \Delta A - \bm X^T \bm \gamma)^2].
\end{align}
The assumptions listed in Assumption \ref{assump:data}, as well as the following bounded variance assumption, are critical for identification of the ATE using the population minimizers in \eqref{dstar} and \eqref{dstar-cov}.
\begin{assumption}[Bounded Variance]\label{assump:variance}
    Define $\tilde \epsilon = Y - \theta^* - \Delta^* A$ and $\epsilon = Y - \theta^*_C - \Delta^*_C A - \bm X^T \bm \gamma^*_C$. Also define $\tilde{\sigma}^2 = \Var(\tilde{\epsilon})$ and $\sigma^2 = \Var(\epsilon)$. We assume that $\tilde{\sigma}^2 < \infty$ and $\sigma^2 < \infty$.
\end{assumption}
The sample analogs of \eqref{dstar} and \eqref{dstar-cov} are given in the following definition.
\begin{definition}[ANOVA and ANCOVA Estimators]\label{def:estimators}
Consider $P$ from Assumption \ref{assump:data}. Suppose that we observe $n^*_k$ data pairs $(y_i, a_i, \bm x_i) \overset{iid}{\sim} P$, $i = 1, .., n^*_k$, with $\bm x_i = (x_{i,1}, ..., x_{i,p})^T$, and where each $\bm x_{\cdot, p}$ has been centered at its mean. We define $\hat{\Delta}_k$ to satisfy the following:
    \begin{align}\label{anova-samp}
    (\hat{\Delta}_k, \hat{\theta}_k) &= \argmin_{\Delta, \theta \in \mathbb{R}}  \sum_{i=1}^{n^*_k} (y_i - \theta - \Delta a_i)^2
    \end{align}
We refer to $\hat{\Delta}_k$ as the ANOVA estimator. Similarly, we define $\hat{\Delta}^C_k$ to satisfy the following:
    \begin{align}\label{ancova-samp}
    (\hat{\Delta}^C_k, \hat{\theta}^C_k, \hat{\bm \gamma}^C_k) &= \argmin_{\Delta, \theta \in \mathbb{R}, \gamma \in \mathbb{R}^p} \sum_{i=1}^{n^*_k} (y_i - \theta - \Delta a_i - \bm x_i^T \bm \gamma)^2.
\end{align}
We refer to $\hat{\Delta}_k^C$ as the ANCOVA estimator. The ANCOVA and ANOVA estimators are simply ordinary least squares estimators of the treatment effect with and without adjusting for the covariates $\bm x_i$, respectively.
\end{definition}

The following lemma summarizes well-known results about ANOVA and ANCOVA in randomized trials \citep{yangEfficiencyStudyEstimators2001, colantuoniLeveragingPrognosticBaseline2015, wangAnalysisCovarianceRandomized2019}.
\begin{lemma}[Identification and Estimation]\label{lemma:identification-estimation}
    Consider $\Delta^{*}$ and $\Delta^{*}_C$ from Definition \ref{def:ate} and $\hat{\Delta}_k$ and $\hat{\Delta}_k^C$ from Definition \ref{def:estimators}. Under Assumptions \ref{assump:data} and \ref{assump:variance},
    \begin{enumerate}[label=(\roman*)]
        \item \textit{Identification}. $\Delta^{*} = \Delta^{*}_C$, and that $2\Delta^* = \E_P[Y|A = 1] - \E_P[Y|A=-1]$.
        \item \textit{Consistency}. $\hat{\Delta}_k$ and $\hat{\Delta}_k^C$ are $\sqrt{n^*_k}$-consistent for $\Delta^{*}$.
        \item \textit{Asymptotic Variance}. $\hat{\Delta}_k$ and $\hat{\Delta}_k^C$ have asymptotic variances $\tilde{\sigma}^2$, and ${\sigma}^2$, respectively. Also, $\sigma^2 \leq \tilde{\sigma}^2$.
    \end{enumerate}
\end{lemma}
Lemma \ref{lemma:identification-estimation}(i) tells us that the population minimizers $\Delta^*$ and $\Delta^{*C}$ in the un-adjusted and adjusted problem are equivalent, and that they are a scalar multiple of the average treatment effect. Lemma \ref{lemma:identification-estimation}(ii) tells us that under standard assumptions for randomized trials, both the ANOVA and ANCOVA estimators can be used to estimate the average treatment effect. Finally, Lemma \ref{lemma:identification-estimation}(iii) confirms that, asymptotically, we will not lose efficiency by performing covariate adjustment. These are classical results \citep{yangEfficiencyStudyEstimators2001, colantuoniLeveragingPrognosticBaseline2015, wangAnalysisCovarianceRandomized2019}. We build off of these critical results in the next section to propose valid methods for interim monitoring, estimation, and inference in the hybrid design. 

\section{Proposed Methods for the Hybrid Design}\label{hybrid}

We now present our methodological contribution. Our work builds off of the results referenced in Section \ref{meth:intro}. Our proposal is presented in three parts.  In the first part of our proposal (Section \ref{sec:joint-dist}), we derive the joint distribution of the ANOVA and ANCOVA estimators and their associated test statistics across stages $1, ..., K$ of a group sequential design. We formally define the monitoring procedure for the hybrid design in terms of a sequence of decisions based on the test statistics that we introduce. In the second part (Section \ref{sec:monitoring}), we use the joint distribution to form interim monitoring boundaries that have correct type I error in the hybrid design. In the third part (Section \ref{sec:estimation-inference}), we discuss how to use the joint distribution to form median unbiased point estimates and asymptotically correct confidence intervals in the hybrid design.

\subsection{Joint distribution of ANOVA and ANCOVA estimators in a group sequential design}\label{sec:joint-dist}

We show that ANOVA and ANCOVA test statistics have a joint asymptotic Gaussian distribution across interim analyses. Additionally, their covariance matrix can be expressed as a relatively simple function of the information fraction (i.e., the relative fraction of observations accrued between two analyses) and the prognostic value of the covariates used in ANCOVA. These results are summarized in Theorem \ref{joint-estimates} and Corollary \ref{joint-statistics}. These multivariate Gaussian distributions are leveraged to form corrected rejection boundaries, point estimates, and confidence intervals in the hybrid design.

\begin{theorem}[Joint Distribution of Estimates]\label{joint-estimates} Consider the asymptotic regime in Assumption \ref{assump:sequential}, and define $\hat{\Delta}_k$ and $\hat{\Delta}_k^C$ from Definition \ref{def:estimators}. Let 
$$\hat{\bm \Delta} := \Big(\hat{\Delta}_1, \hat{\Delta}^{C}_1, ..., \hat{\Delta}_K, \hat{\Delta}^{C}_K \Big)^T$$ and $\bm \Delta^* = \Delta^* \bm 1_{2K}$. In other words, $\hat{\bm \Delta}$ are stacked effect estimates across stages using ANOVA and ANCOVA, and $\bm \Delta^*$ is the corresponding population quantity from equation \eqref{dstar}. Then under Assumptions \ref{assump:data} and \ref{assump:variance},
\begin{align*}
	\sqrt{n} (\hat{\bm \Delta} - \bm \Delta^*) \td N\Big(\bm 0, \bm \Sigma_{\hat{\bm \Delta}} \Big), \quad \bm \Sigma_{\hat{\bm \Delta}} = \bm T \otimes \begin{pmatrix} \tilde{\sigma}^2 & \sigma^2 \\ \sigma^2 & \sigma^2 \end{pmatrix}.
\end{align*}
where $\sigma^2$ and $\tilde{\sigma}^2$ are defined in Assumption \ref{assump:variance} and  $\bm T$ is the matrix with entries $\bm T_{(k,k')}=1/t^{*}_{\max(k,k')}$.

\end{theorem}
The result of Theorem \ref{joint-estimates} is critical for deriving the joint distribution of the test statistics. For hypothesis testing in the hybrid design, we will use the following test statistics:
\begin{definition}[ANOVA and ANCOVA Test Statistics]\label{def:test-stat}
    Consider $(\hat{\Delta}_k, \hat{\theta}_k)$ and $(\hat{\Delta}^C_k, \hat{\theta}^C_k, \hat{\bm \gamma}^C_k)$ as the empirical minimizers from Definition \ref{def:estimators}. Let
    \begin{align*}
        \hat{\sigma}_k = \left[\frac{1}{n_k^*} \sum_{i=1}^{n_k^*} (y_i - \hat{\theta}_k - \hat{\Delta}_k a_i)^2\right]^{1/2} \quad\text{and}\quad \hat{\sigma}^C_k = \left[\frac{1}{n_k^*} \sum_{i=1}^{n_k^*} (y_i - \hat{\theta}^C_k - \hat{\Delta}^C_k a_i - \bm x_i^T \hat{\bm \gamma}^C_k)^2\right]^{1/2}.
    \end{align*}
    We define $Z_k^* = \sqrt{n_k^*} \hat{\Delta}_k / \hat{\sigma}_k$ and $Z_k^{*C} = \sqrt{n_k^*} \hat{\Delta}_{k}^C / \hat{\sigma}_k^C$ as the ANOVA and ANCOVA test statistics at stage $k$.
\end{definition}
 
\begin{corollary}[Joint Distribution of Test Statistics]\label{joint-statistics}
Consider the definitions of the ANOVA and ANCOVA test statistics $Z_k^*$ and $Z_k^{*C}$ from Definition \ref{def:test-stat}. Let $\bm Z^* = (Z^*_1, Z^{*C}_1, ..., Z_K^*, Z^{*C}_K)^T$ be the vector of stacked test statistics. Then under Assumptions \ref{assump:data} and \ref{assump:variance}, and the null hypothesis $\Delta^{*} = 0$,
\begin{align*}
	\bm Z^* \Rightarrow N\left(\bm 0, \bm \Sigma_{\bm Z^*}\right) \quad \bm \Sigma_{\bm Z^*} = \bm T' \otimes \begin{psmallmatrix}
1 & \rho \\
\rho & 1
\end{psmallmatrix}
\end{align*}
where $\rho = \sigma / \tilde{\sigma}$ from Assumption \ref{assump:variance}. Finally, let $\bm T'$ be the  matrix with entries $\bm T'_{(k,k')} = \sqrt{t^*_{\min(k, k')}/t^*_{\max(k, k')}}$.
\end{corollary}
Therefore, the joint distribution of the ANOVA and ANCOVA test statistics across stages of a group sequential trial is a function of the information fractions and the ratio of the asymptotic variances of ANOVA and ANCOVA. Theorem \eqref{joint-estimates} and Corollary \ref{joint-statistics} are the primary results that we use to derive monitoring boundaries, adjusted estimates, and confidence intervals for the hybrid design.
 \begin{procedure}[Hybrid Design of a Group Sequential Trial]\label{def:hybrid} 
 The hybrid design is as follows.
    \begin{enumerate}
        \item Select an asymptotically acceptable sequence $\{u_k; k = 1, ..., K\}$ of critical values (discussed in corollary COR) for a two-sided hypothesis test using the test statistics in Definition \ref{def:test-stat}.
        \item For $k = 1, ..., K$:
        \begin{enumerate}
            \item Enroll $n_k$ participants and randomize with equal probability.
            \item Test to see if we should stop at the current stage and reject the null. If $k < K$, check if $|Z_k^*| \geq u_k$. If $k = K$, check if $|Z_K^{*C}| \geq u_K$.
        \end{enumerate}
        \item After termination of the trial due to early stopping or continuing through stage $K$, construct a point estimate and confidence interval that takes into account both covariate adjustment and the group sequential design (see section SEC and appendix APP for details).
    \end{enumerate}
\end{procedure}

For clarity of exposition (and to match what is most likely done in practice), we focus on the hybrid design in Definition \ref{def:hybrid} and illustrated the design in Figure \ref{fig:gst}(b). However, we note that as long as we have the joint distribution of test statistics from \eqref{joint-statistics}, we are not restricted to only performing covariate adjustment at the end of a group sequential trial. We are also not restricted to having symmetric (or two-sided) monitoring boundaries. Everything that we discuss in the following sections can be easily modified to account for an arbitrary sequence of un-adjusted and covariate-adjusted estimators across stages of a trial, asymmetric monitoring boundaries, and one-sided monitoring boundaries.

\subsection{Interim monitoring in the hybrid design}\label{sec:monitoring}

Our goal in this section is to derive a stopping boundary $\{u_k; k = 1, ..., K\}$ such that we control the overall probability of making a type I error. In particular, we want to find a $\{u_k\}$ that satisfies Definition \ref{def:critical}, given below. We emphasize that we do not propose a new, ``hybrid'' estimator to preserve type I error. Rather, we derive monitoring boundaries that preserve type I error when the well-known ANOVA and ANCOVA estimators from Definition~\ref{def:estimators} and their test statistics from Definition~\ref{def:test-stat} are used in an hybrid trial design described in Definition \ref{def:hybrid}.
\begin{definition}[Asymptotically Acceptable Sequence of Critical Values]\label{def:critical}
    Consider the hybrid design in Definition \ref{def:hybrid}. Let $P_0$ be the joint distribution of test statistics $(Z_1^*, ..., Z_{K-1}^*, Z_K^{*C})$ under the null hypothesis that $\Delta^* = 0$. We call $\{u_k; k = 1, ..., K\}$ an \emph{asymptotically acceptable sequence of critical values} if
    \begin{align}\label{eq:type-1-error}
        \lim_{n\to\infty} P_0(|Z_1^*| \geq u_1 \text{ or } |Z_{2}^*| \geq u_2 \text{ or } ... \text{ or } |Z_{K-1}^*| \geq u_{K-1} \text{ or } |Z_K^{*C}| \geq u_K) = \alpha
    \end{align}
    where $\alpha$ is a pre-specified type I error probability.
\end{definition}
We have already derived the joint asymptotic distribution of the test statistics in Corollary \ref{joint-statistics}. In addition to the information fractions, knowledge (or an accurate estimate) of $\rho \equiv \sigma / \tilde{\sigma}$ is the key quantity that determines the joint distribution. There are infinite choices of asymptotically acceptable sequences $\{u_k\}$. We focus on the O'Brien-Fleming (OBF) \citep{obrienMultipleTestingProcedure1979} style of boundary, but note that these can be modified for any shape that is desired (e.g., for Pocock \citep{pocockGroupSequentialMethods1977}, the shape has $c$ for each stage, rather than $c/\sqrt{k}$). We present two examples of asymptotically acceptable sequences of critical values for the hybrid design in Corollaries \ref{corr:unif-inflated} and \ref{corr:end-inflated}.
\begin{corollary}[Uniform-Inflated OBF Critical Values]\label{corr:unif-inflated}
    Let $\bm U^{\rho} \sim N_K(\bm 0, \bm \Sigma_{\rho})$ with
    \begin{align}\label{eq:sigma-rho}
        \bm \Sigma(\rho) := \begin{pmatrix}
            1 & \sqrt{t^*_1/t^*_2} & ... & \sqrt{t^*_1/t^*_{K-1}} & \rho \sqrt{t^*_1/t^*_K} \\
            \sqrt{t^*_1/t^*_2} & ... & ... & ... & \rho \sqrt{t^*_2/t^*_K} \\
            ... & ... & ... & ... & ... \\
            \sqrt{t^*_1/t^*_{K-1}} & ... & ... & ... & \rho \sqrt{t^*_{K-1}/t^*_{K}} \\
            \rho \sqrt{t^*_1/t^*_{K}} & \rho \sqrt{t^*_2/t^*_K} & ... & \rho \sqrt{t^*_{K-1}/t^*_{K}} & 1
        \end{pmatrix}
    \end{align}
    Under Assumptions \ref{assump:sequential}, \ref{assump:data}, and \ref{assump:variance}, define $c$ as the value that satisfies the following:
    \begin{align}\label{eq:obf-critical-value}
        P(|U_1^{\rho}| \geq c \text{ or } |U_{2}^{\rho}| \geq c / \sqrt{2} \text{ or } ... \text{ or } |U_{K-1}^{\rho}| \geq c / \sqrt{K-1} \text{ or } |U_K^{\rho}| \geq c/\sqrt{K}) = \alpha.
    \end{align}
    Then $\{c, c / \sqrt{2}, ..., c / \sqrt{K}\}$ is an asymptotically acceptable sequence of critical values for $(Z_1^*, ..., Z_{K-1}^*, Z_K^{*C})$ in the hybrid design described in Procedure \ref{def:hybrid}. 
\end{corollary}
\begin{corollary}[End-Inflated OBF Critical Values]\label{corr:end-inflated}
    Let $\bm U^{\rho} \sim N_K(\bm 0, \bm \Sigma(\rho))$ and $\bm U^1 \sim N_K(\bm 0, \bm \Sigma(1))$ with $\bm \Sigma$ defined in Corollary \ref{corr:unif-inflated}. Note, by $\bm \Sigma(1)$ we mean to replace all values of $\rho$ in \eqref{eq:sigma-rho} with $1$.
    Define $c$ as the value that satisfies
    \begin{align}\label{eq:obf-critical-value-2}
        P(|U_1^{1}| \geq c \text{ or } |U_{2}^{1}| \geq c / \sqrt{2} \text{ or } ... \text{ or } |U_{K-1}^{1}| \geq c / \sqrt{K-1} \text{ or } |U_K^{1}| \geq c/\sqrt{K}) = \alpha.
    \end{align}
    Subsequently, define $c'$ as the value that satisfies
    \begin{align}\label{eq:obf-critical-value-3}
        P(|U_1^{\rho}| \geq c \text{ or } |U_{2}^{\rho}| \geq c / \sqrt{2} \text{ or } ... \text{ or } |U_{K-1}^{\rho}| \geq c / \sqrt{K-1} \text{ or } |U_K^{\rho}| \geq c'/\sqrt{K}) = \alpha.
    \end{align}
    % Further, let $\alpha^*$ be defined as $$\alpha^* := P(|U_1^1| \geq c \text{ or } ... \text{ or } |U_{K-1}^1| \geq c / \sqrt{K-1})$$ i.e., $\alpha^*$ is the cumulative type I error spent by stage $K - 1$. Define $c'$ as the value that satisfies the following
    % \begin{align*}
    %     P(|U_K^{\rho}| \geq c'/\sqrt{K} \text{ and } |U_{K-1}^{\rho}| < c/\sqrt{K-1}, ..., |U_1^{\rho}| < c) &= \alpha - \alpha^*.
    % \end{align*}
    Then $\{c, c/\sqrt{2}, ..., c/\sqrt{K-1}, c'/\sqrt{K}\}$ is an asymptotically acceptable sequence of critical values for $(Z_1^*, ..., Z_{K-1}^*, Z^{*C}_K)$ in the hybrid design described in Procedure \ref{def:hybrid}.
\end{corollary}
The critical values above do not have a closed form, but can be calculated numerically without too much difficulty (see Supplemental Material B.2). The critical values in Corollary \ref{corr:unif-inflated} have the exact same shape as the classical OBF boundaries (Figure 2), but are uniformly inflated to account for the decreased correlation between the last stage test statistic with previous stages. In contrast, the critical values in Corollary \ref{corr:end-inflated} follow the classical OBF boundaries up to the last stage. At the last stage of the trial, the critical value is inflated to a level that ensures exactly $\alpha$ type I error is spent (Figure 3). 
In practice, it is unlikely that we will know $\rho$ during trial design. This makes the critical values in Corollary \ref{corr:unif-inflated} less practical. However, the boundaries in Corollary \ref{corr:end-inflated} can still be used as long as we have a consistent estimator of $\rho$ at the end of the trial.

We have focused on the hybrid design where covariate adjustment is performed when the trial ends (Procedure \ref{def:hybrid}). However, our methods are readily extended to the setting where covariate adjustment is performed at any stage(s) of a trial. In this case, we recommend using ``alpha-spending'' \citep{lanDiscreteSequentialBoundaries1983}: rather than pre-specifying the critical values at each stage, they are calculated on the fly based on a pre-specified amount of type I error to be spent at each stage. We provide some further details about how one might employ alpha-spending in the Supplementary Material.

\subsection{Point estimates, $p$-values, and confidence intervals in the hybrid design}\label{sec:estimation-inference}

Once the trial has ended, we want to compute point estimates and confidence intervals for the treatment effect and $p$-values for the test of the null hypothesis. Computing these quantities without taking into consideration the group sequential design may lead to incorrect inference: point estimates may be biased \citep{kimPointEstimationFollowing1989}, and inference based on $p$-values and confidence intervals may be anti-conservative \citep{kimConfidenceIntervalsFollowing1987}. Methods have been proposed to adjust point estimates (median unbiased point estimates), $p$-values, and confidence intervals based on the joint distribution of the test statistics under the null and alternative hypotheses \citep{kimPointEstimationFollowing1989, kimConfidenceIntervalsFollowing1987, siegmundEstimationFollowingSequential1978, gernotGroupSequentialConfirmatory2016}. We use the same approach, but with the joint distribution of the test statistics and effect estimates from Corollary \ref{joint-statistics} and Theorem \ref{joint-estimates}. 

These methods rely on an ordering of test statistics in the sample space. Group sequential trials have a two-dimensional test statistic: the observed statistic and the stage at which it was observed. Sample space orderings of this two-dimensional test statistic are more nuanced in the hybrid design. Furthermore, computation of p-values based on these orderings is more complicated in the hybrid design. Importantly, our methodology can be applied in scenarios where both un-adjusted and covariate-adjusted analyses are performed within the same stage of a trial after early trial termination (often adjustment will be performed once the trial is stopped and the covariate values are collected).
In the Supplemental Material, we give full details of our strategy for forming adjusted estimates, p-values, and confidence intervals using the key results from Theorem \ref{joint-estimates} and Corollary \ref{joint-statistics}.

\section{Results}

We analyze the performance of our adjustments for interim monitoring boundaries, point estimates, and confidence intervals via simulation. We base the trial designs on two published trials in cystic fibrosis (CF).

\subsection{Simulation parameters}

The GROW \citep{bozicOralGlutathioneGrowth2020} trial was an RCT of 58 children with CF between 2-10 years old to evaluate the effect of oral glutathione on their growth. The primary outcome was change in weight-for-age Z-score from visits 2-4. There was one interim look after 25 participants (about 50\% accrual). The trial protocol specified ANOVA at the interim analysis and ANCOVA at the final analysis, using the weight-for-age z-score at visit 3 as a covariate. The TEACH \citep{nicholsTestingEffectsCombining2021} trial was an RCT of 108 people over 12 years old with CF and chronic Pseudomonas aeruginosa infection. The goal was to evaluate the effect of adding oral azithromycin to inhaled tobramycin on lung function. The trial protocol specified ANOVA at two interim analyses (after 50\% and 75\% accrual) and a final analysis with ANCOVA using the randomization strata as a covariate. We explore the use of additional baseline covariates available for analysis in the TEACH and GROW trials as well.

Based on the GROW and TEACH trials, we consider sample sizes of 50, 100, and 250, and 1000 for a 3-stage trial with information fractions $0.33$, $0.67$, and $1.0$. We fix $\tilde{\sigma}^2 = 1$, and $\Var[X_j] = 1$, $j = 1, ..., p$, and consider $p = 1$. Given a $\rho$, it's then possible to solve for $\gamma$ in $\bm \gamma = \gamma \bm 1$, i.e., all of the covariates having the same coefficient (see Supplemental Material). We consider effect sizes for $\Delta$ of 0.0, 0.1, 0.2, and 0.5, prognostic values of baseline covariates $\rho$ of 0.25, and 0.5 (smaller values of $\rho$ corresponding to higher prognostic value). We present results for Pocock-type boundaries. See Supplemental Material for 2-stage trials, two covariates, and OBF-type boundaries.

\subsection{Type I error and power under hybrid designs with proposed modifications to interim monitoring boundaries}

We analyze type I error and power in a simulation study considering five scenarios of how a trial might be conducted, outlined below. Results are shown in Figure 4.
\begin{enumerate}[label=(\Alph*)]
	\item The same analysis method is performed for interim monitoring and the final analysis, and critical values are computed during trial design.
	\begin{enumerate}[label=(\roman*)]
		\item ANOVA or (ii) ANCOVA is used for both monitoring and final analysis.
	\end{enumerate}
	\item Hybrid design: ANOVA is used for interim monitoring, ANCOVA is used in the final analysis
	\begin{enumerate}[label=(\roman*)]
		\item No adjustment of the monitoring boundaries.
		\item $\rho$ is known and is used to compute critical values during trial design (Figure 2).
		\item $\rho$ is estimated at the last stage of the trial and the last critical value is inflated accordingly (Figure 3).
	\end{enumerate}
\end{enumerate}
At small sample sizes, the type I error is inflated for all trial procedures. As the sample size increases, most procedures have nominal type I error. Importantly, the type I error when switching from ANOVA to ANCOVA without adjusting the boundaries is uniformly high. Both types of adjustment (uniform- and end-inflated) correct this inflation in type I error. The implication is that boundaries need to be adjusted in the hybrid design, especially when highly prognostic covariates are used. In the Supplemental Material we explore the impact of incorrectly specifying $\rho$ at the beginning of the trial when computing boundaries.

Figure 5 shows the power corresponding to these different trial procedures. Expectedly, more prognostic covariates increases the power of procedures that use covariate adjustment. Importantly, there is not a large difference between the power in the hybrid design scenarios. The implication is that adjusting the boundaries to control type I error in the hybrid design does not hurt power.

\subsection{Performance of adjusted point estimates and confidence intervals in the hybrid design}

In Table 1, we show bias and coverage of several strategies to create point estimates and confidence intervals in the hybrid design. The first strategy is to compute point estimates and CIs ignoring the group sequential design (``simple''). The second strategy takes into account the group sequential design using stage-wise ordering, but not the hybrid design (``GS''). The third strategy takes into account the group sequential hybrid design with our modified stage-wise ordering (``GS + Adjust'').

The simple point estimate and confidence interval generally has small median bias and acceptable coverage. However, there are several scenarios where ``GS + Adjust'' substantially out performs has the simple point estimate (e.g., $\Delta = 0.1$, $\rho = 0.5$, $n = 250, 1000$). In almost all scenarios, ``GS + Adjust'' has higher coverage than the simple confidence intervals. The point estimates for ``GS + Adjust'' are constructed so that they will be median unbiased; as a result, there is a small amount of mean bias in the point estimates.

Importantly, ignoring the hybrid design when using the group sequential adjustment for point estimates and confidence intervals has serious consequences. Under some simulation parameters, the ``GS'' approach leads to large bias, particularly for covariate adjustment using highly prognostic covariates. The confidence intervals may have over- or under-coverage. In some cases, the under-coverage is extreme (e.g., 16\%). The implication of these simulation results is that if point estimates and confidence intervals use standard group sequential adjustment methods, they should be modified for the hybrid design.

\section{Conclusions}

Although it would be ideal to conduct covariate adjustment during interim monitoring \citep{vanlanckerImprovingInterimDecisions2019a}, investigators may not have the entire set of prognostic baseline covariates ready to use. Standard group sequential trial methodology relies on the same hypothesis testing procedure being conducted at each stage of the trial. Failing to account for the hybrid design may result in incorrect decisions on rejection of the null hypothesis, and may lead to incorrect point estimates and confidence intervals. Inconsistent use of covariate adjustment changes the covariance between group sequential test statistics. We derived the covariance matrix of the test statistics in a group sequential trial under the hybrid design. We used it to construct group sequential monitoring boundaries that preserve type I error, and to obtain correct estimation and inference for treatment effects under a slightly modified stage-wise ordering of the test statistics.

Our simulations demonstrate two key implications of inconsistent covariate adjustment. If baseline covariates are expected to be prognostic, but will not be used until the end of the trial, then it is important to adjust the monitoring boundaries for the hybrid design. In some cases, failing to adjust the boundaries resulted in a 1-2\% inflation in type I error, even at large sample sizes. A small inflation in type I error can have a big impact on the Bayesian posterior probability of a beneficial treatment \citep{emersonBayesianEvaluationGroup2007}. Modifying the boundaries using our strategies brings the type I error to the nominal level. Additionally, if group sequential adjustment for point estimates and confidence intervals is desired, these should further be adjusted for the hybrid design. Ignoring the hybrid design can result in  high bias and extremely low coverage.

Though we focused on a particular trial design, the methods that we discuss here may be extended to other designs. We focused on two-arm trials with simple randomization and equal probability of being randomized to each arm. Similar derivations could be used for trials that use unequal randomization probabilities and trials with more than one treatment arm. Our approach could be extended to other covariate-adjusted estimators that are asymptotically linear (as others \citep{vanlanckerCombiningCovariateAdjustment2022a} have noted), since the covariance matrix for the hybrid design can be characterized using the estimators' influence functions.

Constructing monitoring boundaries and performing group sequential-adjusted estimation and inference requires computing probabilities of regions of multivariate Gaussians. Standard group sequential methodology uses recursive integration techniques which simplify the calculation of probabilities of a multivariate Gaussian to a sequence of probabilities from univariate Gaussians \citep{armitageRepeatedSignificanceTests1969}. Recursive integration relies on ``independent increments,'' i.e., that incremental changes to test statistics across stages in a trial are independent \citep{kimIndependentIncrementsGroup2020}. The hybrid design breaks the independent increments property. Therefore, rather than use recursive integration techniques, we use numerical integration across regions of a multivariate Gaussian. Using standard statistical software, this integration is feasible for trials that have a reasonable number of stages (e.g., $\leq 5$). Others have pursued strategies that orthogonalize estimators across consecutive stages in order to recover independent increments and avoid integrating multivariate Gaussians \citep{vanlanckerCombiningCovariateAdjustment2022a}. However, it is not necessary for the settings we have discussed in this paper.

Finally, future work may consider extensions to the covariate-adaptive randomization setting. Using a working model for covariate adjustment that includes treatment-by-covariate interactions achieves a guaranteed efficiency gain and universal applicability under a variety of covariate-adaptive randomization strategies where the ANCOVA procedure we have discussed here does not \citep{yeBetterPracticeCovariate2022}. Extensions of our work to the covariate-adaptive randomization setting would be particularly useful for trials where categorical covariates are used for randomization, but where estimates adjusted for additional clinically relevant baseline covariates are desired at the end of a trial.

\section{Conflict of Interest}
The Authors declare declare that there is no conflict of interest.

\section{Funding}
SLH is funded by the US Cystic Fibrosis Foundation (Grant: HELTSHE18Y3) and National Institutes of Health NIDDK (Grant: P30 DK089507). NS is supported by the US Cystic Fibrosis Foundation (Grant: RAMSEY03Y0).

\section{Supplemental Materials}
Supplemental materials that contain technical proofs and results are available online. The code to reproduce simulations and make the tables and figures is available at: \url{https://github.com/mbannick/RCTCovarAdjust/releases/tag/resubmission} \citep{marlena_bannick_2022_7036090}.

\bibliographystyle{apalike}
\bibliography{refs}

\newpage
\begin{figure}
	\caption{Uniform-inflated Pocock and OBF bounds in a 4-stage trial with equally-sized stages.}\label{bounds-adjusted}
	\includegraphics[width=\textwidth]{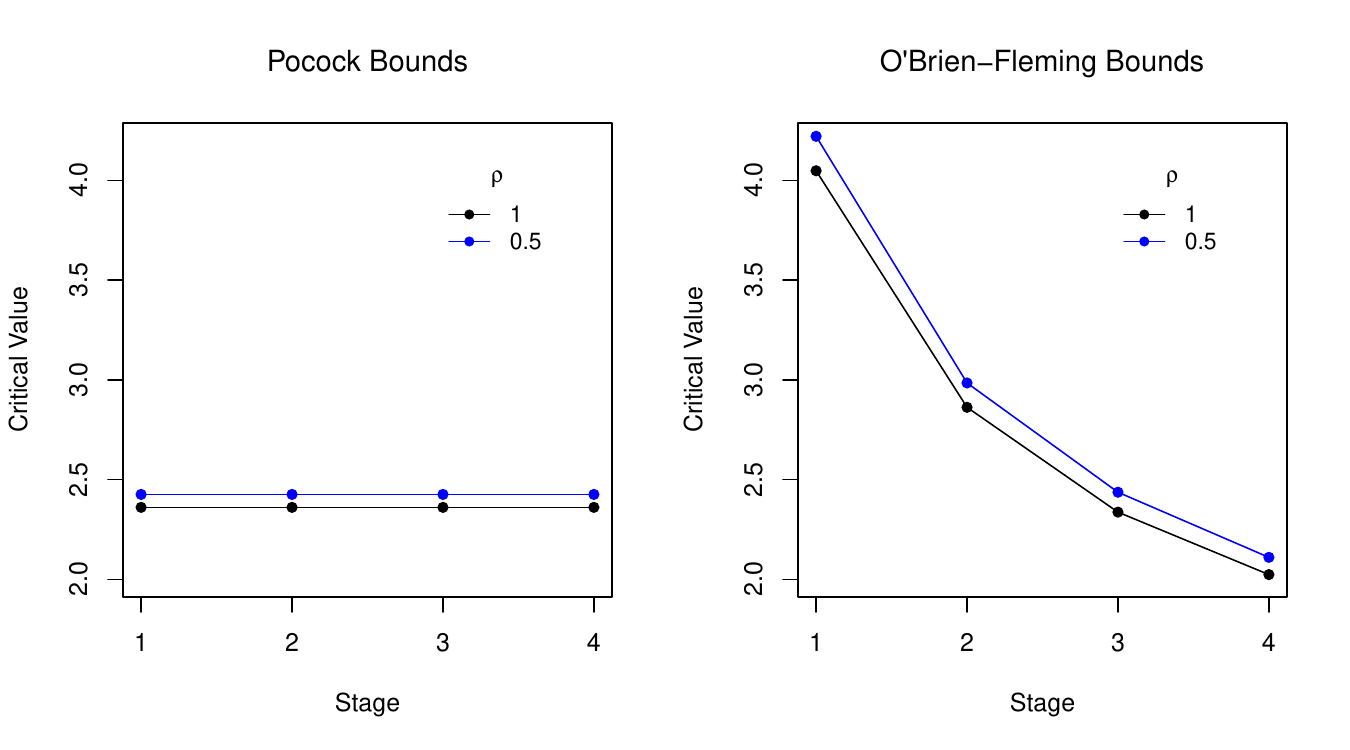}
	\caption{End-inflated Pocock and OBF bounds in a 4-stage trial with equally-sized stages.}\label{bounds-adjusted-alpha}
	\includegraphics[width=\textwidth]{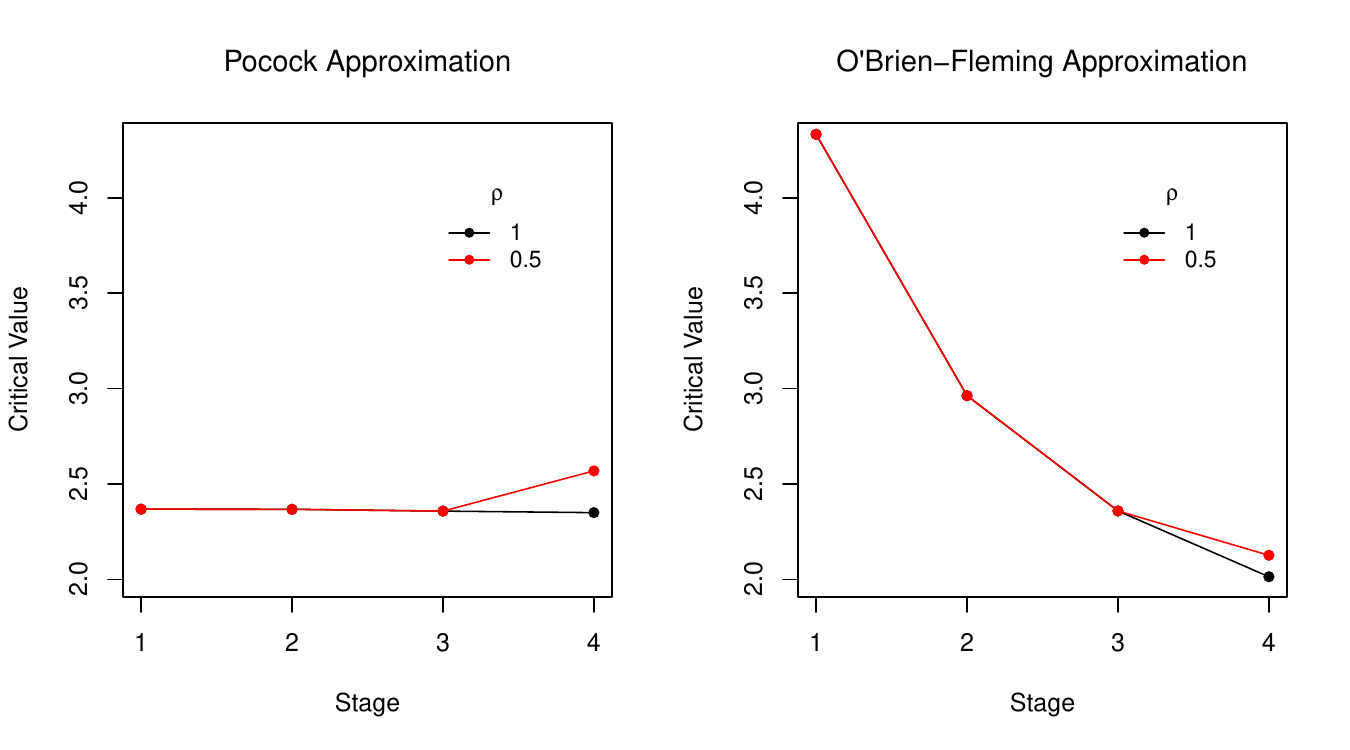}
\end{figure}
\begin{figure}[htbp!]
	\caption{Type I error under scenarios (A)(i)-(ii) and (B)(i)-(iii) across 25,000 simulated trials.}\label{fig:type1error}
	\includegraphics[width=\textwidth]{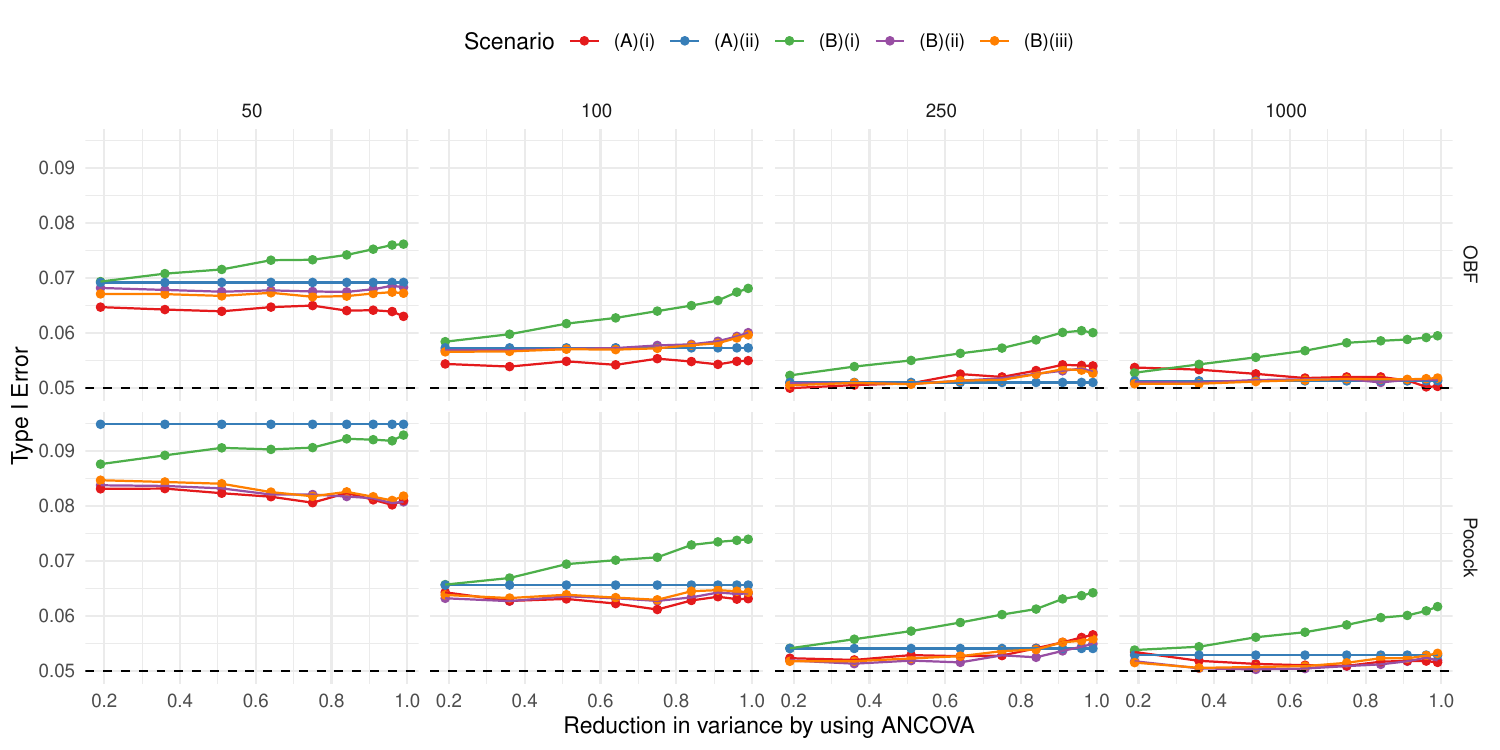}
	\caption{Power under scenarios (A)(i)-(ii) and (B)(i)-(iii) with an effect size of 0.1 across 25,000 simulated trials.}\label{fig:power}
	\includegraphics[width=\textwidth]{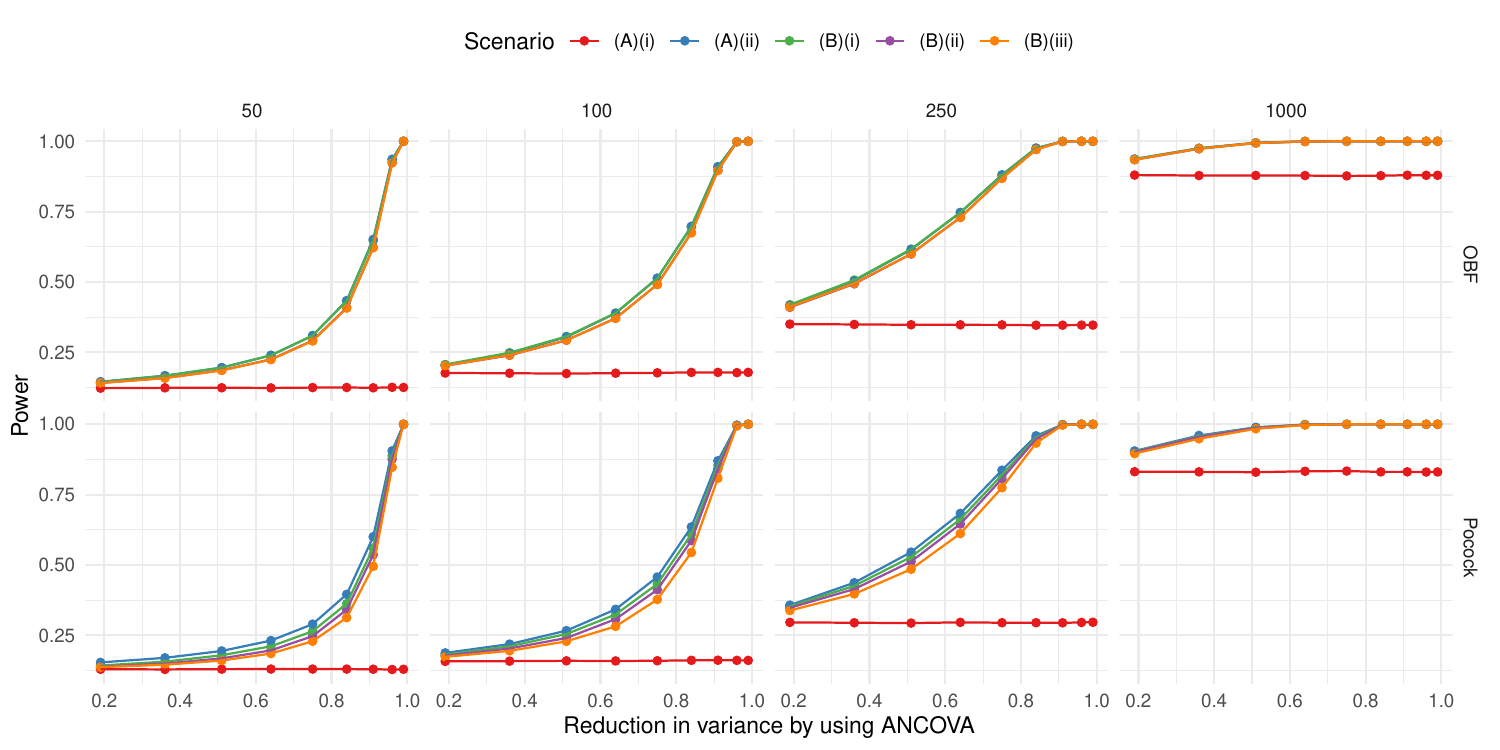}
\end{figure}
% latex table generated in R 4.1.2 by xtable 1.8-4 package
% Mon Apr 25 17:47:59 2022
\begin{table}[htbp!]
\centering
\caption{Bias ($\times 100$) and coverage results based on 10,000 simulated trials for each row, end-inflated Pocock boundaries for 3-stage trial. Bias is shown as median bias and mean bias in parentheses. $\Delta$ is the true effect size, $\rho$ is the true ratio of standard error between the ANCOVA and ANOVA estimators, and $n$ is the total sample size for the trial.}\label{tab:bias-cover}
\begin{tabular}{ccc|ccc|ccc}
  \hline
  \hline
& & & \multicolumn{3}{c|}{Scaled Bias: Median (Mean)} & \multicolumn{3}{c}{Coverage} \\
\hline
$\Delta$ & $\rho$ & $n$ & Simple & GS & GS + Adjust & Simple & GS & GS + Adjust \\ 
  \hline
  0.00 & 0.25 &  50 & -0.03 (0.01) & -0.14 (-0.16) & -0.04 (0.03) & 0.93 & 0.94 & 0.92 \\ 
  0.00 & 0.25 & 100 & 0.01 (-0.00) & -0.05 (-0.11) & 0.02 (-0.03) & 0.94 & 0.95 & 0.94 \\ 
  0.00 & 0.25 & 250 & 0.02 (0.01) & -0.02 (-0.07) & 0.02 (0.05) & 0.95 & 0.96 & 0.94 \\ 
  0.00 & 0.25 & 1000 & -0.00 (-0.00) & -0.02 (-0.04) & 0.00 (0.02) & 0.95 & 0.96 & 0.95 \\ 
   \hline 
  0.00 & 0.50 &  50 & -0.05 (0.03) & -0.21 (-0.26) & -0.06 (-0.02) & 0.92 & 0.94 & 0.92 \\ 
  0.00 & 0.50 & 100 & 0.02 (-0.01) & -0.11 (-0.23) & 0.03 (0.01) & 0.93 & 0.95 & 0.94 \\ 
  0.00 & 0.50 & 250 & 0.03 (0.01) & -0.02 (-0.11) & 0.03 (0.01) & 0.94 & 0.96 & 0.95 \\ 
  0.00 & 0.50 & 1000 & -0.00 (-0.01) & -0.03 (-0.07) & 0.00 (-0.01) & 0.95 & 0.96 & 0.95 \\ 
   \hline 
  0.10 & 0.25 &  50 & 0.03 (0.15) & -1.63 (-1.95) & 0.06 (1.99) & 0.93 & 0.94 & 0.93 \\ 
  0.10 & 0.25 & 100 & 0.07 (0.09) & -3.38 (-3.16) & 0.12 (1.87) & 0.94 & 0.96 & 0.94 \\ 
  0.10 & 0.25 & 250 & 0.06 (0.09) & -5.68 (-4.92) & 0.01 (1.58) & 0.95 & 0.18 & 0.94 \\ 
  0.10 & 0.25 & 1000 & 0.03 (0.09) & -6.87 (-4.90) & -0.02 (0.35) & 0.95 & 0.30 & 0.95 \\ 
   \hline 
  0.10 & 0.50 &  50 & 0.15 (0.58) & -0.30 (-0.75) & 0.10 (1.78) & 0.92 & 0.93 & 0.93 \\ 
  0.10 & 0.50 & 100 & 0.19 (0.41) & -0.39 (-1.08) & 0.16 (1.56) & 0.93 & 0.94 & 0.94 \\ 
  0.10 & 0.50 & 250 & 0.21 (0.40) & -1.85 (-1.73) & 0.05 (1.35) & 0.94 & 0.95 & 0.95 \\ 
  0.10 & 0.50 & 1000 & 0.16 (0.36) & -3.78 (-2.78) & -0.01 (0.47) & 0.95 & 0.62 & 0.95 \\ 
   \hline 
  0.20 & 0.25 &  50 & 0.09 (0.26) & -10.24 (-8.74) & 0.20 (3.56) & 0.93 & 0.29 & 0.94 \\ 
  0.20 & 0.25 & 100 & 0.10 (0.24) & -12.97 (-10.45) & 0.17 (2.87) & 0.94 & 0.16 & 0.94 \\ 
  0.20 & 0.25 & 250 & 0.12 (0.19) & -13.80 (-9.74) & -0.05 (0.65) & 0.95 & 0.31 & 0.95 \\ 
  0.20 & 0.25 & 1000 & -0.00 (0.01) & -0.12 (-1.51) & -0.02 (-0.54) & 0.95 & 0.86 & 0.95 \\ 
   \hline 
  0.20 & 0.50 &  50 & 0.39 (1.08) & -3.01 (-2.80) & 0.35 (3.06) & 0.92 & 0.93 & 0.93 \\ 
  0.20 & 0.50 & 100 & 0.43 (0.92) & -6.36 (-4.37) & 0.36 (2.56) & 0.93 & 0.94 & 0.94 \\ 
  0.20 & 0.50 & 250 & 0.43 (0.72) & -7.71 (-5.57) & 0.03 (0.91) & 0.95 & 0.60 & 0.95 \\ 
  0.20 & 0.50 & 1000 & 0.01 (0.08) & -0.15 (-1.06) & -0.03 (-0.34) & 0.95 & 0.87 & 0.95 \\ 
   \hline 
  0.50 & 0.25 &  50 & 0.19 (0.48) & -34.85 (-21.51) & -0.61 (-0.01) & 0.92 & 0.37 & 0.93 \\ 
  0.50 & 0.25 & 100 & 0.15 (0.24) & -1.75 (-11.15) & -0.12 (-1.81) & 0.94 & 0.67 & 0.94 \\ 
  0.50 & 0.25 & 250 & 0.01 (0.02) & -0.02 (-0.54) & 0.01 (-0.25) & 0.95 & 0.94 & 0.95 \\ 
  0.50 & 0.25 & 1000 & -0.02 (-0.01) & -0.02 (-0.01) & -0.02 (-0.01) & 0.95 & 0.95 & 0.95 \\ 
   \hline 
  0.50 & 0.50 &  50 & 0.89 (1.82) & -20.22 (-12.68) & -0.79 (1.04) & 0.92 & 0.56 & 0.92 \\ 
  0.50 & 0.50 & 100 & 0.43 (0.83) & -2.43 (-7.32) & -0.22 (-0.77) & 0.94 & 0.67 & 0.94 \\ 
  0.50 & 0.50 & 250 & 0.03 (0.06) & -0.02 (-0.36) & 0.00 (-0.15) & 0.95 & 0.94 & 0.95 \\ 
  0.50 & 0.50 & 1000 & -0.04 (-0.02) & -0.04 (-0.02) & -0.04 (-0.02) & 0.95 & 0.95 & 0.95 \\ 
   \hline  
 \hline
\end{tabular}
\end{table}

\appendix

\newpage
\doublespacing

\section{Review Material: Covariate Adjustment in Randomized Trials}\label{app:review}

In this section, we re-prove well-known results about ANCOVA in randomized trials with simple randomization and equal treatment allocation \citep{yangEfficiencyStudyEstimators2001, colantuoniLeveragingPrognosticBaseline2015, wangAnalysisCovarianceRandomized2019}. The lemmas in this section form the proof of Lemma \ref{lemma:identification-estimation} for identification and estimation. In Lemma \ref{pop-min} (Lemma \ref{lemma:identification-estimation}(i)) we first show that the population parameter estimated by ANCOVA is directly related to the average treatment effect. In Lemmas \ref{samp-min} and \ref{asymptotic-variance}, we show the asymptotic behavior of ANCOVA and compare that to the behavior of ANOVA (Lemma \ref{lemma:identification-estimation}(ii) and (iii)). Since ANOVA and ANCOVA can be written in the same way as the ordinary least squares (OLS) estimator, we first prove a results about OLS, and then apply them to the ANOVA and ANCOVA setting.

\subsection{Parameter Identification}

\begin{lemma}[Parameter Identification for OLS]\label{ols-identification}

Consider $(\bm Z, Y) \sim P$ where $P$ is the population joint distribution of $\bm Z$ and $Y$, and $\bm Z \in \mathbb{R}^d$. We define
\begin{align*}
    \bm \beta^{*} = \arg\min_{\bm \beta \in \mathbb{R}} \E_P[(Y - \bm Z^T \bm \beta)^2]
\end{align*}
Assume that $\Var(Y - \bm Z^T \bm \beta) < \infty$.
Then $\bm \beta^* = \E_P[\bm Z \bm Z^T]^{-1} \E_P[\bm Z Y]$.
\end{lemma}

\begin{proof}
Note that this loss function is convex in $\bm \beta$, so we can find a unique minimizer $\bm \beta^*$. Thus, $\bm \beta^*$ is the solution to:
\begin{align*}
    \frac{\partial}{\partial \bm \beta} \int (y - \bm z^T \bm \beta)^2 \quad dP(\bm z, y) &= 0 \\
    \int 2 \bm x (y - \bm z^T \bm \beta) \quad dP(\bm z, y) &= 0 \\
    \int 2 \bm z y \quad dP(\bm z, y) &= \int 2 \bm x \bm z^T \bm \beta \quad dP(\bm z, y) \\
    \E_P[\bm Z Y] &= \E_P[\bm Z \bm Z^T] \bm \beta \\
    \bm \beta^* &= \E_P[\bm Z \bm Z^T]^{-1} \E_P[\bm Z Y].
\end{align*}
\end{proof}

\begin{lemma}[Identification of Average Treatment Effect: Lemma \ref{lemma:identification-estimation}(i)]\label{pop-min}
Under Assumptions \ref{assump:data} and \ref{assump:variance}, the population minimizer for \eqref{dstar} is $(\Delta^*, \theta^*) = \E_P[(A, 1)(A, 1)^T] \E_P[(A, 1)^T Y]$, and population minimizer for \eqref{dstar-cov} is $$(\Delta^*_C, \theta^*_{C}, \bm \gamma^{*T}_C)^T = \E_P[(A, 1, \bm X^T)^T(A, 1, \bm X^T)]^{-1} \E_p[(A, 1, \bm X^T) Y].$$ In both cases, because of the independence of $A$ and $\bm X$ from Assumption \ref{assump:data},
\begin{align}\label{ate}
    2\Delta^* = 2\Delta^*_C = \E_P[Y|A = 1] - \E_P[Y|A=-1].
\end{align}
\end{lemma}

\begin{proof}
ANOVA and ANCOVA are special cases of the general least squares formulation that we have presented. Consider the two sets of predictors,
\begin{equation*}
\begin{split}
    \text{ANOVA: } \quad \bm \beta &:= (\Delta, \theta) \\
    \text{ANCOVA: } \quad \bm \beta_C &:= (\Delta_C, \theta_C, \bm \gamma_C)
    \end{split}
    \begin{split}
    \quad \bm Z &:= (A, 1) \\
    \quad \bm Z_C &:= (A, 1, \bm X)
    \end{split}
\end{equation*}
where $\bm X \in \mathbb{R}^p$ are covariates, and $A \in \{-1, 1\}$ is a treatment group indicator. In the ANCOVA model, we assume that $\bm X \indep A$, i.e. covariates are independent of treatment assignment in the population. Additionally, assume that we have equal randomization probabilities for the two treatment groups,
\vspace{-0.75cm}
\begin{singlespace}
\begin{align*}
    P[A = a] = \begin{cases}
    0.5 \quad \mbox{if} \quad a = 1 \\
    0.5 \quad \mbox{if} \quad a = -1.
    \end{cases}
\end{align*}
\end{singlespace}
Using the form of $\bm \Omega$ as defined from Theorem \ref{ols-identification}, let $\bm \Omega = \E_P[\bm Z \bm Z^T]$, and $\bm \Omega_C = \E_P[\bm Z_C \bm Z_C^T]$. Note that $\E[A] = 0$, and $\E[A^2] = 1$ since $A^2 = 1$ with probability 1, and that $\E[A X_j] = 0$ since the covariates are independent of the treatment assignment. Therefore,
\begin{align*}
    \bm \Omega = E_P[(A, 1)(A, 1)^T] &= \begin{pmatrix}
    \E_P[A^2] & \E_P[A] \\
    \E_P[A] & \E_P[1^2]
    \end{pmatrix} = 
    \begin{pmatrix}
    1 & 0 \\
    0 & 1
    \end{pmatrix} \\
    \bm \Omega_C = \E_P[(A, 1, \bm X^T)^T (A, 1, \bm X^T)] &= \begin{pmatrix}
    \bm \Omega_{1,1} & \bm \Omega_{1,2} & \E_P[A \bm X^T] \\
    \bm \Omega_{2,1} & \bm \Omega_{2,2} & \E_P[\bm X^T] \\
    \E_P[A\bm X] & \E_P[\bm X] & \E_P[\bm X \bm X^T]
    \end{pmatrix} =
    \begin{pmatrix}
    1 & 0 & \bm 0_p^T \\
    0 & 1 & \E_P[\bm X^T] \\
    \bm 0_p & \E_P[\bm X] & \E_P[\bm X \bm X^T]
    \end{pmatrix}.
\end{align*}
In each case, $\bm \Omega$, and $\bm \Omega_C$ are block-diagonal, so the inverse of the matrix is the inverse of each block, $\bm \Omega^{-1} = \bm I_2$, and $ \bm \Omega^{-1}_C = \diag{1, \E_P[(1, \bm X^T)^T (1, \bm X^T)]^{-1}}$.
Recall from Theorem \ref{ols-identification} that $\bm \beta^* = \bm \Omega^{-1} \E_P[\bm Z Y]$.
Therefore, for the ANOVA and ANCOVA problem,
\begin{align*}
    \bm \beta^* \equiv (\Delta^*, \theta^*) &= \bm \Omega^{-1} \E_P[(A, 1)^T Y] = \Big\{\E_P[AY], \E_P[Y]\Big\}^T \\
    \bm \beta_C^* \equiv (\Delta^*_C, \theta_C^*, \bm \gamma^*_C) &= \bm \Omega^{-1}_C \E_P[(A, 1, \bm X)^T Y] = \Big\{\E_P[A Y], \E_P[(1, \bm X^T)^T (1, \bm X^T)]^{-1}\E_P[(1, \bm X^T) Y]\Big\}^T.
\end{align*}
Thus, $\Delta^* = \Delta^*_C$, and
\begin{align*}
    \Delta^* &= \E_P[A Y] \\
    &= \E_P[A Y | A = 1] P[A = 1] + \E_P[A Y | A = -1] P[A = -1] \\
    &= \E_P[Y|A = 1] P[A = 1] + \E_P[-Y | A = -1] P[A = -1] \\
    &= \E_P[Y | A = 1] P[A= 1] - \E_P[Y| A = -1] P[A = -1] \\
    &= \frac{1}{2} \Big\{ \E_P[Y|A= 1] - \E_P[Y | A = -1] \Big \}.
\end{align*}
\end{proof}

\subsection{Asymptotic Behavior of ANOVA and ANCOVA}

In this section, we derive the influence functions for ANOVA and ANCOVA to show that they are consistent and asymptotically normal estimators. Furthermore, we use the asymptotic forms to derive their variance. In these lemmas, we will use $n$ as our sample size. In the main text, we instead use $n^*_k$, to be consistent with the group sequential trials notation. By Assumption \ref{assump:sequential}, $n^*_k \to \infty$ as $n \to \infty$, so these results are equivalent.

\begin{definition}[Asymptotically Linear Estimator]\label{asymptotic-linear}
An estimator $\hat{\psi}$ is an asymptotically linear estimator if it can be written as
\begin{align*}
    \hat{\psi}_n - \psi = \frac1n \sum_{i=1}^{n} \phi(o_i) + o_p\Big(n^{-\frac12}\Big) \quad \mathrm{where} \quad \E_P[\phi] = 0.
\end{align*}
We call $\phi$ the influence function for the estimator $\hat{\psi}$, that takes as an argument all of the data for a single observation $o_i$.
\end{definition}

\begin{lemma}[Influence Function for OLS Estimator]\label{influence-ols}
Let $\bm z = (\bm z_1, ..., \bm z_d)$ be a design matrix of covariates and $\bm y = (y_1, ..., y_n)$ be an outcome vector. The multivariate influence function for the least squares estimator $\hat{\bm \beta}$ of $\bm \beta^*$ is $\bm \phi(o_i) = \bm \Omega^{-1} \bm z_i (y_i - \bm z_i^T \bm \beta^*)$ where $\bm \Omega$ is $\E[\bm Z \bm Z^T]$. Furthermore, $\hat{\bm \beta}$ is $\sqrt{n}$-consistent for $\bm \beta^*$, defined in Theorem \ref{ols-identification}.
\end{lemma}

\begin{proof}
The least squares estimator for $\bm \beta^*$ is $\hat{\bm \beta} = (\bm z^T \bm z)^{-1} \bm z \bm y$, since it minimizes:
\begin{align*}
    \hat{\bm \beta} = \argmin_{\beta \in \mathbb{R}^p} \sum_{i=1}^{n} (y_i - \bm z_i^T \bm \beta^*)^2.
\end{align*}
We can see this by rewriting the above in matrix form and taking the gradient with respect to $\bm \beta$:
\begin{align*}
    \frac{\partial}{\partial \bm \beta} ||\bm y - \bm z \bm \beta ||^2 = 2 \bm z^T (\bm y - \bm z \bm \beta) &= 0 \\
    \bm z^T \bm z \bm \beta &= \bm z^T \bm y \\
    \hat{\bm \beta} &= (\bm z^T \bm z)^{-1} \bm z^T \bm y.
\end{align*}
Define $\E[\bm Z \bm Z^T] = \bm \Omega$. Our goal is to find the multivariate influence function for $\hat{\bm \beta} - \bm \beta^*$. Define $\epsilon = Y - \bm Z^T \bm \beta$, or equivalently $Y = \bm Z \bm \beta + \epsilon$. Let $\bm \epsilon = (\epsilon_1, ..., \epsilon_n)$ where $\epsilon_i = y_i - \bm z_i^T \bm \beta$. Using the form of $\hat{\bm \beta}$, we have
\begin{align*}
    \hat{\bm \beta} &= (\bm z^T \bm z)^{-1} \bm z^T \bm y \\
    &= (\bm z^T \bm z)^{-1} \bm z^T (\bm z\bm \beta^* + \bm \epsilon) \\
    &= (\bm z^T \bm z)^{-1} \bm z^T \bm z \bm \beta^* + (\bm z^T \bm z)^{-1} \bm z^T \bm \epsilon \\
    &= \bm \beta^* + (\bm z^T \bm z)^{-1} \bm z^T \bm \epsilon \\
    &\implies \hat{\bm \beta} - \bm \beta^* = (\bm z^T \bm z)^{-1} \bm z^T \bm \epsilon.
\end{align*}
To derive the influence function, we want to find the right-hand side in terms of a sum of independent observations plus some second order term(s). We notice that $\bm z^T \bm z$ is the empirical second moment of $\bm Z$, multiplied by $n$. We can rewrite the above in this form, and then add and subtract the true second moment to get it in terms of the population parameters:
\begin{align*}
    \hat{\bm \beta} - \bm \beta^* &= \frac{1}{n} \Bigg(\frac{\bm z^T \bm z}{n}\Bigg)^{-1} \bm z^T \bm \epsilon \\
    &= \frac{1}{n} \Bigg(\frac{\bm z^T \bm z}{n}\Bigg)^{-1} \sum_{i=1}^n \bm z_i \epsilon_i \\
    &= \frac{1}{n} \Bigg(\frac{\bm z^T \bm z}{n}\Bigg)^{-1} \sum_{i=1}^n \bm z_i \epsilon_i + \frac{1}{n} \bm \Omega^{-1} \sum_{i=1}^n \bm z_i \epsilon_i - \frac{1}{n} \bm \Omega^{-1} \sum_{i=1}^n \bm z_i \epsilon_i \\
    &= \frac{1}{n} \bm \Omega^{-1} \sum_{i=1}^n \bm z_i \epsilon_i + \Bigg\{ \Bigg(\frac{\bm z^T \bm z}{n}\Bigg)^{-1} - \bm \Omega^{-1} \Bigg\} \frac{1}{n} \sum_{i=1}^n \bm z_i \epsilon_i.
\end{align*}
Focusing on the right-most term, we will show that
\begin{align}\label{ols-sec}
    \Bigg\{ \Bigg(\frac{\bm z^T \bm z}{n}\Bigg)^{-1} - \bm \Omega^{-1} \Bigg\} \frac{1}{n} \sum_{i=1}^n \bm z_i \epsilon_i = o_p(n^{-\frac{1}{2}}).
\end{align}
By the WLLN, $\frac{1}{n} \sum_{i=1}^{n} \bm z_i \epsilon_i \to \E_P[\bm Z \epsilon]$ in probability. However,
\begin{align*}
    \E_P[\bm Z \epsilon]
    &= \E_P[\bm Z Y - \bm Z \bm Z^T \bm \beta^*] \\
    &= \E_P[\bm Z Y] - \E_P[\bm Z \bm Z^T] \bm \beta^* \\
    &= \E_P[\bm Z Y] - \E_P[\bm Z \bm Z^T] \E_P[\bm Z \bm Z^T]^{-1} \E_P[\bm Z Y] \\
    &= \E_P[\bm Z Y] - \E_P[\bm Z Y] = \bm 0.
\end{align*}
Thus, by the central limit theorem, $\sqrt{n} \frac{1}{n} \sum_{i=1}^{n} \bm z_i \epsilon_i = O_P(1)$, which implies that $\frac{1}{n} \sum_{i=1}^{n} \bm z_i \epsilon_i = O_P(n^{-\frac12})$. By the WLLN, $\frac{\bm z^T \bm z}{n} = \frac{1}{n} \sum_{i=1}^{n} \bm z_i \bm z_i^T \to \E[\bm Z \bm Z^T] = \bm \Omega$, which again by the central limit theorem means $(\frac{\bm z^T \bm z}{n} - \bm \Omega) = O_p(n^{-\frac12})$. Applying the continuous mapping theorem by noting that the inverse of a matrix is a continuous function, $((\bm z^T \bm z / n)^{-1} - \bm \Omega^{-1}) = O_P(n^{-\frac12})$. Using these two parts, \eqref{ols-sec} is $O_p(n^{-\frac{1}{2}}) O_p(n^{-\frac{1}{2}}) = O_p(n^{-1}) = o_P(n^{-\frac12})$, which means \eqref{ols-sec} is a second-order term, and can be ignored asymptotically. Further, note that $\E[\bm \phi] = 0$, since $\E[\bm \phi] = \bm \Omega^{-1} \E[\bm Z \epsilon]$, which we have shown above to be $\bm 0$. Therefore, the multivariate influence function for $\hat{\bm \beta}$ is
\begin{align}\label{multivariate-influence}
    \bm \phi(o_i) = \bm \Omega^{-1} \bm z_i \epsilon_i = \bm \Omega^{-1} \bm z_i (y_i - \bm z_i^T \bm \beta^*).
\end{align}
Since $\hat{\bm \beta} - \bm \beta^* = \frac{1}{n} \sum_{i=1}^{n} \bm \phi(o_i) + o_P(n^{-\frac12})$, we have $\hat{\bm \beta} \tp \bm \beta^*$ in probability (by the WLLN since $\E[\bm \phi] = 0$, and applying Slutsky's Theorem), i.e., $\hat{\bm \beta} \tp \bm \beta^* = \E_P[\bm Z \bm Z^T]^{-1} \E_P[\bm Z Y]$ (Lemma \ref{ols-identification}).
\end{proof}

\begin{lemma}[Consistency of ANOVA and ANCOVA Estimators: Lemma \ref{lemma:identification-estimation}(ii)]\label{samp-min}
Denote $\bm x = (\bm x_1, ..., \bm x_n)$, $\bm a = (a_1, ..., a_n)^T$, and $\bm y = (y_1, ..., y_n)^T$. Define $\epsilon_i = y_i - \theta^*_C - \Delta^*_C a_i - \bm x_i^T \bm \gamma^*_C$ and $\tilde{\epsilon}_i = y_i - \theta^* - \Delta^* a_i$ as the sample versions of $\epsilon$ and $\tilde{\epsilon}$. Under Assumptions \ref{assump:data} and \ref{assump:variance}, the sample minimizers of \eqref{anova-samp} and \eqref{ancova-samp} are 
\begin{align*}
    (\hat{\Delta}, \hat{\theta}) &= ((\bm a, \bm 1)^T(\bm a, \bm 1))^{-1} (\bm a, \bm 1)^T \bm y \\
    (\hat{\Delta}_C, \hat{\theta}_C, \hat{\bm \gamma}_C) &= ((\bm a, \bm 1, \bm x)^T (\bm a, \bm 1, \bm x))^{-1} (\bm a, \bm 1, \bm x)^T \bm y,
\end{align*}
$\hat{\Delta}$ and $\hat{\Delta}_C$ are asymptotically linear estimators with influence functions $\phi(o_i) := a_i(\epsilon_i + \theta^*_C - \theta^* + \bm x_i^T \bm \gamma^*_C)$ and $\phi_C(o_i) := a_i \epsilon_i$, and $\hat{\Delta}$ and $\hat{\Delta}_C$ are $\sqrt{n}$-consistent estimators for $\Delta^*$.
\end{lemma}

\begin{proof}
We directly apply the results from Lemma \ref{influence-ols}, using the formulations in the previous proof of Lemma \ref{pop-min}. Using the definitions of the ANOVA and ANCOVA problem in the Proof of Lemma \ref{pop-min}, and the definition of the OLS estimator from Lemma \ref{influence-ols}, the sample minimizers in the ANOVA and ANCOVA problems are given by
\begin{align*}
    (\hat{\Delta}, \hat{\theta}) &= ((\bm a, \bm 1)^T(\bm a, \bm 1))^{-1} (\bm a, \bm 1)^T \bm y \\
    (\hat{\Delta}_C, \hat{\theta}_C, \hat{\bm \gamma}_C) &= ((\bm a, \bm 1, \bm x)^T (\bm a, \bm 1, \bm x))^{-1} (\bm a, \bm 1, \bm x)^T \bm y,
\end{align*}
By Lemma \ref{influence-ols}, $\bm \phi(o_i) = \bm \Omega^{-1} \bm z_i \epsilon_i$ where $\epsilon_i$ is the error term for the generic least squares problem. In the ANCOVA problem, we have $\epsilon_i = y_i - \theta^*_C - \Delta^*_C a_i - \bm x_i^T \bm \gamma^*_C$. In the ANOVA problem, we can rewrite the error term $\tilde{\epsilon}_i$:
\begin{align*}
    \tilde{\epsilon}_i - \epsilon_i = (y_i - y_i) + (\theta^*_C - \theta^*) + (\Delta^*_C - \Delta^*) a_i + \bm x_i^T \bm \gamma^*_C.
\end{align*}
However, by Lemma \ref{pop-min}, we know that $\Delta^*_C = \Delta^*$. So we have $\tilde{\epsilon}_i = \epsilon_i + (\theta^*_C - \theta^*) + \bm x_i^T \bm \gamma^*_C$. Then the multivariate influence functions have the form:
\begin{align*}
    \bm \phi(o_i) &= \bm \Omega^{-1} (a_i, 1)^T \tilde{\epsilon}_i = \bm \Omega^{-1} (a_i, 1)^T (\epsilon_i + \theta^*_C - \theta^* + \bm x^T_i \bm \gamma^*_C) \\
    \bm \phi^C(o_i) &= \bm \Omega^{-1}_C (a_i, 1, \bm x_i^T)^T \epsilon_i
\end{align*}
and the influence functions for just the $\Delta^*$ component of the regression parameters, corresponding to the $\bm \Omega_{1,1}^{-1} = 1$ component, are $\phi(o_i) = a_i (\epsilon_i + \theta^*_C - \theta^* + \bm x_i^T \bm \gamma^*_C)$ and $\phi^C(o_i) = a_i \epsilon_i$. Finally, by Lemma \ref{influence-ols}, $\hat{\Delta}_C$ and $\hat{\Delta}$ are both $\sqrt{n}$-consistent for $\Delta^*$.
\end{proof}

\begin{lemma}[Asymptotic Variance of ANOVA and ANCOVA Estimators: Lemma \ref{lemma:identification-estimation}(iii)]\label{asymptotic-variance}
Define $\sigma^2 = \Var[\epsilon]$ where $\epsilon = Y - \theta^*_C - \Delta^*_C A - \bm X^T \bm \gamma^*_C$ as in Assumption \ref{assump:variance}. Then under Assumptions \ref{assump:data} and \ref{assump:variance}, $\Var[(\phi, \phi_C)^T] = \begin{psmallmatrix} \tilde{\sigma}^2 & \sigma^2 \\ \sigma^2 & \sigma^2 \end{psmallmatrix}$ where $\tilde{\sigma}^2 = \sigma^2 + \Var[(\bm X^T \bm \gamma^*_C)^2]$.
\end{lemma}

\begin{proof}
We calculate the asymptotic variance of the ANOVA and ANCOVA estimators based on their influence functions from Lemma \ref{samp-min}. As in Lemma \ref{samp-min}, define the random variables $\epsilon = Y - \theta^*_C - A\Delta^*_C - \bm X^T \bm \gamma_C^*$ and $\tilde{\epsilon} = Y - \theta^* - A \Delta^*$. Then by the same logic as Lemma \ref{samp-min}, we have $\tilde{\epsilon} - \epsilon = (\theta^*_C - \theta^*) + \bm X^T \bm \gamma_C^*$. By definition of the population minimizer, $(\theta^*_C, \Delta^*_C, \gamma^*_C)$ and $(\theta^*, \Delta^*)$ solve the population score equations for ANCOVA and ANOVA, respectively. The score equations have expectation zero, hence, $\E[(1, A, \bm X^T)^T(Y - \theta^*_C - A\Delta^*_C - \bm X^T \bm \gamma^*_C)] = \bm 0$ and $\E[(1, A)^T (Y - \theta^* - A\Delta^*) ] = \bm 0$. Let $\Var[\epsilon] = \sigma^2$. Then,
\begin{align*}
    \Var[\phi_C] &= \Var[A\epsilon] \\
    &= \E[A^2 \epsilon^2] - \E[A(Y - \theta_C^* - \Delta^*_C A - \bm X^T \bm \gamma^*_C)]^2 \\
    &= \E[\epsilon^2] = \Var[\epsilon] = \sigma^2 \\
    \Var[\phi] &= \Var[A(\theta_C^* - \theta^* + \bm X^T \bm \gamma^*_C + \epsilon)] \\
    &= \Var[A\epsilon] + \Var[A(\theta_C^* - \theta^* + \bm X^T \bm \gamma^*_C)] + 2\Cov[A\epsilon, A(\theta_C^* - \theta^* + \bm X^T \bm \gamma^*_C)] \\
    &= \sigma^2 + \Var[\E(A(\theta_C^* - \theta^* + \bm X^T \bm \gamma^*_C)|A)] + \E[\Var(A(\theta_C^* - \theta^* + \bm X^T \bm \gamma^*_C)|A)] + 2\Cov[A\epsilon, A(\theta_C^* - \theta^* + \bm X^T \bm \gamma^*_C)] \\
    &= \sigma^2 + \Var[A(\E(\theta_C^* - \theta^* + \bm X^T \bm \gamma^*_C)] + \E[\Var(\theta_C^* - \theta^* + \bm X^T \bm \gamma^*_C)] + 2\Cov[A\epsilon, A(\theta_C^* - \theta^* + \bm X^T \bm \gamma^*_C)] \\
    &= \sigma^2 + \E[\theta_C^* - \theta^* + \bm X^T \bm \gamma^*_C]^2 + \Var[\theta_C^* - \theta^* + \bm X^T \bm \gamma^*_C] + 2\Cov[A\epsilon, A(\theta_C^* - \theta^* + \bm X^T \bm \gamma^*_C)] \\
    &= \sigma^2 + \E[\tilde{\epsilon} - \epsilon]^2 + \Var[\bm X^T \bm \gamma^*_C] + 2\Cov[A\epsilon, A(\theta_C^* - \theta^* + \bm X^T \bm \gamma^*_C)] \\
    &= \sigma^2 + \Var[\bm X^T \bm \gamma^*_C] + 2\left\{\E[A^2 \epsilon (\theta^*_C - \theta^* + \bm X^T \bm \gamma^*_C)] - \E[A \epsilon] \E[A(\theta^*_C - \theta^* + \bm X^T \bm \gamma^*_C)] \right\} \\
    &= \sigma^2 + \Var[\bm X^T \bm \gamma^*_C] + 2\E[\epsilon (\theta^*_C - \theta^* + \bm X^T \bm \gamma^*_C)] \\
    &= \sigma^2 + \Var[\bm X^T \bm \gamma^*_C] + 2(\bm \gamma^*_C)^T \E[\bm X \epsilon] \\
    &= \sigma^2 + \Var[\bm X^T \bm \gamma^*_C] \\
    \Cov[\phi, \phi^C] &= \Cov[A \epsilon, A (\bm X^T \bm \gamma + \epsilon)] \\
    &= \E[A^2 \epsilon (\bm X^T \bm \gamma + \epsilon)] - \E[A \epsilon] \E[A (Y - \theta_C^* - \Delta^*_C A)] \\
    &= \E[\epsilon \bm X^T \bm \gamma_C^*] + \E[\epsilon^2] \\
    &= (\bm \gamma_C^{*})^T \E[\bm X(Y - \theta^*_C - \Delta^*_C A - \bm X^T \bm \gamma_C^*)] + \Var[\epsilon] \\
    &= \Var[\epsilon] = \sigma^2
\end{align*}
where we have repeatedly applied the relationships from the score equations.
\end{proof}
Let $\rho^2 = \sigma^2 / \tilde{\sigma}^2$, i.e., $\rho^2$ is the ratio of variance of ANCOVA to ANOVA. If $\rho = 1$, then $\Var[(\bm X^T \bm \gamma^*_C)^2]$ must be 0, i.e., there is no relationship between the covariates and the outcome.
\begin{remark}\label{pop-min-rho}
The asymptotic variance of the ANCOVA estimator is no larger than the asymptotic variance of the ANOVA estimator $(0 < \rho \leq 1)$, and the reduction in variance by using ANCOVA is $\Var[(\bm X^T \bm \gamma^*_C)^2]$. The reduction in variance by using ANCOVA is a function of how predictive the covariates $\bm x$ are with respect to the outcome $y$.
\end{remark}

\section{The Hybrid Design}\label{app:statistics}

We use the results from the previous review lemmas in Section \ref{app:review} on covariate adjustment to show our desired results for the hybrid design.

\subsection{Joint Distribution of ANOVA and ANCOVA Test Statistics in Group Sequential Design}

\begin{lemma}\label{influence-vcov-multivariate}
Let $n^*_k = \sum_{k'=1}^{k} n_{k'}$, the sample size accumulated up through stage $k$, with $n = n^*_K$, as defined in Assumption \ref{assump:sequential}. Consider the influence functions for ANOVA and ANCOVA $\phi$ and $\phi^C$ from Lemma \ref{samp-min}, and let 
\begin{align*}
    \bar{\phi}^*_k &:= \frac{1}{n^*_k} \sum_{i=1}^{n^*_k} \phi(o_i) \equiv \frac{1}{n^*_k} \sum_{i=1}^{n^*_k} a_i(\bm x_i^T \bm \gamma^*_C + \theta^*_C - \theta^* + \epsilon_i) \\
    \bar{\phi}^{*C}_k &:= \frac{1}{n^*_k} \sum_{i=1}^{n^*_k} \phi^C(o_i) \equiv \frac{1}{n^*_k} \sum_{i=1}^{n^*_k} a_i \epsilon_i.
\end{align*}
Then under Assumptions \ref{assump:data} and \ref{assump:variance},
\begin{align*}
    \Cov[\sqrt{n} \bar{\phi}^*_k, \sqrt{n} \bar{\phi}^*_{k'}] &= t^{*-1}_{\max(k,k')} \tilde{\sigma}^2 \\
    \Cov[\sqrt{n} \bar{\phi}^{*C}_k, \sqrt{n} \bar{\phi}^*_{k'}]
    = \Cov[\sqrt{n} \bar{\phi}^{*C}_k, \sqrt{n} \bar{\phi}^{*C}_{k'}] &= t^{*-1}_{\max(k,k')} \sigma^2
\end{align*}

\end{lemma}
\begin{proof}[Proof of Lemma \ref{influence-vcov-multivariate}]
We make use of the variance of the influence functions that we derived in Lemma \ref{asymptotic-variance}. Recall that $\Var[\phi] = \tilde{\sigma}^2$ and $\Var[\phi^C] = \sigma^2$. Then,
\begin{align*}
    \Cov[\sqrt{n} \bar{\phi}^*_k, \sqrt{n} \bar{\phi}^{*C}_k] 
    &= n \Cov\Bigg[\frac{1}{n_k^*} \sum_{i=1}^{n_k^*} \phi(o_i),  \frac{1}{n_{k'}} \sum_{i=1}^{n^*_{k'}} \phi^C(o_i) \Bigg] \\
    &= \frac{n}{n_k^* n^*_{k'}} \sum_{i=1}^{n_k^*} \sum_{j=1}^{n^*_{k'}} \Cov[\phi(o_i), \phi^C(o_j)] \\
    &= \frac{n}{n_k^* n_{k'}^*} \sum_{i=1}^{\min(n_k^*, n^*_{k'})} \Cov[\phi(o_i), \phi^C(o_i)] \\
    &= \frac{n}{\max(n_k^*, n^*_{k'})} \Cov[\phi(o_i), \phi^C(o_i)]
    = t_{\max(k, k')}^{*-1} \sigma^2.
\end{align*}
Similarly,
\begin{align*}
    \Cov[\sqrt{n} \bar{\phi}^*_k, \sqrt{n} \bar{\phi}^*_k]
    &= \frac{n}{\max(n_k^*, n^*_{k'})} \Cov[\phi(o_i), \phi(o_i)] = t_{\max(k, k')}^{*-1} \tilde{\sigma}^2 \\
    \Cov[\sqrt{n} \bar{\phi}^{*C}_k, \sqrt{n} \bar{\phi}^{*C}_k]
    &= \frac{n}{\max(n_k^*, n^*_{k'})} \Cov[\phi^C(o_i), \phi^C(o_i)]
    = t_{\max(k, k')}^{*-1} \sigma^2.
\end{align*}
\end{proof}

\begin{lemma}[Asymptotic Joint Distribution of Empirical Averages of Influence Functions]\label{joint-influence}
Consider $\bar{\phi}^*_{k}$ and $\bar{\phi}^{*C}_{k}$ from Lemma \ref{influence-vcov-multivariate}. Define $\bar{\bm \phi}^* := (\bar{\phi}^*_{1}, \bar{\phi}^{*C}_{1} ..., \bar{\phi}^*_{K}, \bar{\phi}^{*C}_{K})^T \in \mathbb{R}^{2K}$. Then under Assumptions \ref{assump:sequential}-\ref{assump:variance}, $\sqrt{n} \bar{\bm \phi}^*$ converges in distribution to a multivariate normal, i.e.,
\begin{align*}
    \sqrt{n} \bar{\bm \phi}^* \Rightarrow N(0, \bm \Sigma_{\bar{\bm \phi}^*}) \quad\quad \bm \Sigma_{\bar{\bm \phi}^*} = \bm T \otimes \Var[(\phi, \phi^C)^T]
\end{align*}
 where $\bm T_{k,k'} = t^{*-1}_{\max(k,k')}$.
\end{lemma}

\begin{proof}
Define $\bar{\bm \phi} = (\bar{\phi}_1, \bar{\phi}_1^{C}, ..., \bar{\phi}_K, \bar{\phi}_K^{C}) \in \mathbb{R}^{2K}$. This is the vector of the empirical averages of the influence functions for only the observations obtained in each stage (as opposed to the cumulative vector $\bar{\bm \phi}^*$). Let $n_0 = 0$. We will start by showing that (i) $\sqrt{n} \bar{\bm \phi} \Rightarrow N(0, \bm \Sigma_{\bar{\bm \phi}})$, and (ii) then will multiply it by a matrix in order to get the cumulative vector $\sqrt{n} \bar{\bm \phi}^*$.

Starting with (i), for each element of $\bar{\bm \phi}$,
\begin{align*}
    \sqrt{n} \bar{\phi}_k &= \sqrt{n} \frac{1}{n_k} \sum_{i=n_{(k-1)}+1}^{n_k} \phi(o_i) \\
    &= \frac{1}{\sqrt{n_k}} \sum_{i=n_{(k-1)}+1}^{n_k} t_k^{-1/2} \phi(o_i).
\end{align*}
Similarly, $\sqrt{n} \bar{\phi}^C_k = \frac{1}{\sqrt{n_k}} \sum_{i=n_{(k-1)}+1}^{n_k} t_k^{-1/2} \phi^C(o_i)$. By the CLT, we know that
\begin{align*}
    \sqrt{n}(\phi_k, \phi_k^C) &\Rightarrow N\Bigg(\bm 0, \begin{pmatrix}
    \Var(t_{k}^{-1/2} \phi) & \Cov(t_{k}^{-1/2} \phi, t_k^{-1/2} \phi^C) \\
    \Cov(t_{k}^{-1/2} \phi, t_k^{-1/2} \phi^C) & \Var(t_{k}^{-1/2} \phi^C)
    \end{pmatrix} \Bigg) \\
    &= N\Bigg(\bm 0, \begin{pmatrix}
    t_k^{-1} \tilde{\sigma}^2 & t_k^{-1} \sigma^2 \\
    t_k^{-1} \sigma^2 & t_k^{-1} \sigma^2
    \end{pmatrix} \Bigg) \\
    &= N\Big(\bm 0, t^{-1}_k \Var[(\phi, \phi^C)^T]\Big).
\end{align*}
By the Cramer-Wold Theorem, $\sqrt{n} \bar{\bm \phi} \Rightarrow N(0, \bm \Sigma_{\bar{\bm \phi}})$ if for all $\bm a \in \mathbb{R}^{2K}$, $\bm a^T \sqrt{n} \bar{\bm \phi} \Rightarrow N(0, \bm a^T \bm \Sigma_{\bar{\bm \phi}} \bm a)$. Let $\bm b, \bm c \in \mathbb{R}^K$, and $\bm a = (\bm a_1^T, ..., \bm a_K^T)^T = (b_1, c_1, ..., b_K, c_K)^T$ with $\bm a_k = (b_k, c_k)^T, $. Since each of the $(\phi_k, \phi^C_k) \indep (\phi_{k'}, \phi_{k'}^C)$ when $k \neq k'$ (since they share no observations), the covariance matrix $\bm \Sigma_{\bar{\bm \phi}}$ is
\begin{align*}
    \bm \Sigma_{\bar{\bm \phi}} = \diag{t^{-1}_1, ..., t^{-1}_K} \otimes \Var[(\phi, \phi^C)^T],
\end{align*}
i.e., $\bm \Sigma_{\bar{\bm \phi}}$ is a block-diagonal matrix with the $k^{th}$ $2\times 2$ diagonal block equal to $t^{-1}_k \Var[(\phi, \phi^C)^T]$. Our goal is to show that $\bm a^T \sqrt{n} \bar{\bm \phi}$ converges in distribution to a univariate normal distribution with covariance matrix $\bm a^T \bm \Sigma_{\bar{\bm \phi}} \bm a$, which is
\begin{align*}
    \bm a^T \bm \Sigma_{\bar{\bm \phi}} \bm a &= (\bm a_1^T, ..., \bm a_K^T) \diag{t_1^{-1} \Var[(\phi, \phi^C)^T], ..., t_K^{-1} \Var[(\phi, \phi^C)^T]} (\bm a_1^T, ..., \bm a_K^T)^T \\
    &= \sum_{k=1}^{K} t_{k}^{-1} \bm a_k^T \Var[(\phi, \phi^C)^T] \bm a_k \\
    &= \sum_{k=1}^{K} t_k^{-1} \big(b_k^2 \tilde{\sigma}^2 + c_k^2 \sigma^2 + 2 b_k c_k \sigma^2 \big).
\end{align*}
To show that $\bm a^T \sqrt{n} \bar{\bm \phi}$ converges in distribution to a univariate normal distribution, we note that
\begin{align*}
    \bm a^T \sqrt{n} \bar{\bm \phi} &= \bm a_1^T \sqrt{n} (\bar{\phi}_1, \bar{\phi}^C_1)^T, ...,  \bm a_K^T \sqrt{n} (\bar{\phi}_K, \bar{\phi}^C_K)^T \\
    &= \sum_{k=1}^{K} \Big\{b_k \frac{1}{\sqrt{n_k}} \sum_{i=n_{(k-1)}+1}^{n_k} t_k^{-1/2} \phi(o_i) + c_k \frac{1}{\sqrt{n_k}} \sum_{i=n_{(k-1)}+1}^{n_k} t_k^{-1/2} \phi^C(o_i)\Big\} \\
    &= \sum_{k=1}^{K} \Big\{\frac{1}{\sqrt{n_k}} \sum_{i=n_{(k-1)}+1}^{n_k} t_k^{-1/2} \big\{b_k \phi(o_i) + c_k \phi^C(o_i) \big\} \Big\}.
\end{align*}
Define $S_{k,i} = t_k^{-1/2} \big\{b_k \phi(o_i) + c_k \phi^C(o_i) \big\}$ and $S_k = \frac{1}{\sqrt{n_k}} \sum_{i=n_{(k-1)}+1}^{n_k} S_{k,i}$. Now, it is clear that $\bm a^T \sqrt{n} \bar{\bm \phi}$ can be written as a sum of $K$ independent random variables, i.e., $\bm a^T \sqrt{n} \bar{\bm \phi} = \sum_{k=1}^{K} S_k$. Note that $S_{k,i}$, $i = n_{(k-1)}-1, ..., n_k$ are i.i.d. terms, thus, by the Central Limit Theorem,
\begin{align*}
    S_k &= \frac{1}{\sqrt{n_k}} \sum_{i=n_{(k-1)}+1}^{n_k} t_k^{-1/2} \big\{b_k \phi(o_i) + c_k \phi^C(o_i) \big\} \\ &\Rightarrow N(0, \Var[S_{k,i}]) \\
    &= N\Big(0, t_{k}^{-1} \big\{b_k^2 \Var[\phi] + c_k^2 \Var[\phi^C] + 2 b_k c_k \Cov[\phi, \phi^C] \big\} \Big) \\
    &=  N\Big(0, t_k^{-1} \big(b_k^2 \tilde{\sigma}^2 + c_k^2 \sigma^2 + 2 b_k c_k \sigma^2 \big) \Big).
\end{align*}
By Levy's Continuity Theorem (applying characteristic functions), if $S_1, ..., S_K$ are independent random variables, and each $S_k \Rightarrow S'_k \sim N(0, \Var[S_k])$, then $S_1 + \dots + S_K \Rightarrow S' := S'_1 + \dots + S'_K$ with $S' \sim N(0, \sum_{k=1}^{K} \Var[S_k])$. Putting this together with the above,
\begin{align*}
    \bm a^T \sqrt{n} \bar{\bm \phi} &= \sum_{k=1}^{K} S_k \\
    &\Rightarrow N\Big(0, \sum_{k=1}^{K} \Var[S_k] \Big) \\
    &= N\Big(0, \sum_{k=1}^{K} t_k^{-1} \big(b_k^2 \tilde{\sigma}^2 + c_k^2 \sigma^2 + 2 b_k c_k \sigma^2 \big) \Big) \\
    &= N(0, \bm a^T \bm \Sigma_{\bar{\bm \phi}} \bm a).
\end{align*}
Thus, by the Cramer-Wold Theorem, $\sqrt{n} \bar{\bm \phi} \Rightarrow N(0, \bm \Sigma_{\bar{\bm \phi}})$, and this completes (i).

Now for (ii), we apply a linear transformation to $\sqrt{n} \bar{\bm \phi}$ in order to obtain $\sqrt{n} \bar{\bm \phi}^*$. Notice that since $\sqrt{n} \bar{\phi}_k = \frac{\sqrt{n}}{n_k} \sum_{i=n_{(k-1)}+1}^{n_k} \phi(o_i)$ (and similarly for $\bar{\phi}^C_k$),
\begin{align*}
    \sqrt{n} \bar{\phi}^*_k &= \frac{n_1}{n^*_k} \sqrt{n} \bar{\phi}_1 + \dots + \frac{n_k}{n_k^*} \sqrt{n} \bar{\phi}_k \\
    \sqrt{n} \bar{\phi}^{*C}_k &= \frac{n_1}{n^*_k} \sqrt{n} \bar{\phi}^C_1 + \dots + \frac{n_k}{n_k^*} \sqrt{n} \bar{\phi}^C_k.
\end{align*}
Therefore, we can obtain $\sqrt{n}\bar{\bm \phi}^*$ with the linear transformation $\sqrt{n} \bar{\bm \phi}^* = (\bm L \otimes \bm I_2) \sqrt{n} \bar{\bm \phi}$, where $\bm L$ is a lower triangular matrix with entries
\begin{align*}
    \bm L = \begin{pmatrix}
    \frac{n_1}{n^*_1} & 0 & \dots & 0 \\
    \frac{n_1}{n^*_2} & \frac{n_2}{n^*_2} & \dots & 0 \\
    \dots & \dots & \dots & \dots \\
    \frac{n_1}{n^*_K} & \dots & \dots & \frac{n_K}{n_K^*}
    \end{pmatrix}
\end{align*}
We need the Kronecker product with $\bm I_2$ (the $2\times 2$ identity matrix) here because we only want to select the ANOVA entries $\bar{\phi}_k$ for the cumulative $\bar{\phi}^*_k$, and likewise only the ANCOVA entries $\bar{\phi}^C_k$ for the cumulative $\bar{\phi}^{*C}_k$. It will be easier to work with information fractions, so we can write the $(k, k')^{th}$ entry of $\bm L$ (with $k \leq k'$ since it is lower triangular) as $t_k/t_{k'}^*$. By the continuous mapping theorem, we know that 
\begin{align*}
    \sqrt{n}\bar{\bm \phi}^* = (\bm L \otimes \bm I_2) \sqrt{n} \bar{\bm \phi} &\Rightarrow N\big(0, (\bm L \otimes \bm I_2) \bm \Sigma_{\bar{\bm \phi}} (\bm L \otimes \bm I_2)^T\big).
\end{align*}
Define $\bm \Sigma_{\bar{\bm \phi}^*} := (\bm L \otimes \bm I_2) \bm \Sigma_{\bar{\bm \phi}} (\bm L \otimes \bm I_2)^T$. Then, using the matrix product property of the Kronecker product (see Theorem 3 of \citet{zhangKroneckerProductsTheir2013}),
\begin{align*}
    \bm \Sigma_{\bar{\bm \phi}^*} &= (\bm L \otimes \bm I_2) \bm \Sigma_{\bar{\bm \phi}} (\bm L \otimes \bm I_2)\big) \\
    &= (\bm L \otimes \bm I_2)(\diag{t^{-1}_1, ..., t^{-1}_K} \otimes \Var[(\phi, \phi^C)^T])(\bm L \otimes \bm I_2)^T \\
    &= (\bm L \diag{t^{-1}_1, ..., t^{-1}_K} \bm L^T) \otimes (\bm I_2 \Var[(\phi, \phi^C)^T] \bm I_2) \\
    &= (\bm L \diag{t^{-1}_1, ..., t^{-1}_K} \bm L^T) \otimes  \Var[(\phi, \phi^C)^T].
\end{align*}
We will show that $\bm T = (\bm L \diag{t^{-1}_1, ..., t^{-1}_K} \bm L^T)$ where $\bm T_{k,k'} = t^{*-1}_{\max(k,k')}$.
\begin{align*}
	\bm L \diag{t_1^{-1}, ..., t_K^{-1}} \bm L^T &= \begin{pmatrix}
    \frac{n_1}{n^*_1} & 0 & \dots & 0 \\
    \frac{n_1}{n^*_2} & \frac{n_2}{n^*_2} & \dots & 0 \\
    \dots & \dots & \dots & \dots \\
    \frac{n_1}{n^*_K} & \dots & \dots & \frac{n_K}{n_K^*}
    \end{pmatrix}
    \diag{t^{-1}_1, ..., t^{-1}_K} 
    \begin{pmatrix}
    \frac{n_1}{n^*_1} & \frac{n_1}{n^*_2} & \dots & \frac{n_1}{n^*_K} \\
    0 & \frac{n_2}{n^*_2} & \dots & \dots \\
    \dots & \dots & \dots & \dots \\
    0 & 0 & \dots & \frac{n_K}{n_K^*}
    \end{pmatrix} \\
    &= \begin{pmatrix}
    	\frac{n_1}{n^*_1}t_1^{-1} & 0 & \dots & 0 \\
    \frac{n_1}{n^*_2}t_1^{-1} & \frac{n_2}{n^*_2} t_2^{-1} & \dots & 0 \\
    \dots & \dots & \dots & \dots \\
    \frac{n_1}{n^*_K}t_1^{-1} & \dots & \dots & \frac{n_K}{n_K^*} t_K^{-1}
    \end{pmatrix} 
    \begin{pmatrix}
    \frac{n_1}{n^*_1} & \frac{n_1}{n^*_2} & \dots & \frac{n_1}{n^*_K} \\
    0 & \frac{n_2}{n^*_2} & \dots & \dots \\
    \dots & \dots & \dots & \dots \\
    0 & 0 & \dots & \frac{n_K}{n_K^*}
    \end{pmatrix} \\
    &= \begin{pmatrix}
    	\frac{n^2_1}{n^{*2}_1}t_1^{-1} &   \frac{n^2_1}{n^{*}_1 n^{*}_2}t_1^{-1} & \dots & \frac{n^2_1}{n^{*}_1 n^{*}_K}t_1^{-1} \\
    \frac{n^2_1}{n^*_1 n^*_2}t_1^{-1} & \frac{n^2_1 t_1^{-1} + n^2_2 t_2^{-1}}{n^{*2}_2} & \dots & \frac{n^2_1 t_1^{-1} + n^2_2 t_2^{-1}}{n^{*}_2 n^*_K} \\
    \dots & \dots & \dots & \dots \\
    \frac{n^2_1}{n^*_1 n^*_K}t_1^{-1} & \frac{n^2_1 t_1^{-1} + n^2_2 t_2^{-1}}{n_2^* n_K^*} & \dots & \frac{\sum_{k=1}^{K} n_k^{*2} t_k^{-1}}{n_K^{*2}} 
    \end{pmatrix} \\
        &= \begin{pmatrix}
    	\frac{n}{n^{*}_1} & \frac{n}{n^{*}_2} & \dots & \frac{n}{n^{*}_K} \\
    \frac{n}{n^{*}_2} & \frac{n}{n_2^*} & \dots & \frac{n}{n_K^*} \\
    \dots & \dots & \dots & \dots \\
    \frac{n}{n^*_K} & \frac{n}{n_K^*} & \dots & \frac{n}{n_K^*} 
    \end{pmatrix} = 
    \begin{pmatrix}
    	t^{*-1}_{1} & t^{*-1}_{2} & \dots & t^{*-1}_{K} \\
    t^{*-1}_{2} & t^{*-1}_{2} & \dots & t^{*-1}_{K} \\
    \dots & \dots & \dots & \dots \\
    t^{*-1}_{K} & t^{*-1}_{K} & \dots & t^{*-1}_{K} 
    \end{pmatrix} = \bm T,
\end{align*}
where we have used the fact that for $k' \leq k''$, $\frac{\sum_{k=1}^{k'} n_k^{*2} t_k^{-1}}{n_{k'}^{*} n_{k''}^{*}} = \frac{\sum_{k=1}^{k'} n \cdot n_k}{n_{k'}^{*} n_{k''}^{*}} = \frac{n n^*_{k'}}{n_{k'}^{*} n_{k''}^{*}} = \frac{n}{n_{k''}^*}$.
\end{proof}

\begin{proof}[Proof of Theorem \ref{joint-estimates}]
We now prove convergence in distribution for the full influence function, rather than just the first-order approximation. Define $\bar{\bm \phi}^* := (\bar{\phi}^*_{1}, \bar{\phi}^{*C}_{1} ..., \bar{\phi}^*_{K}, \bar{\phi}^{*C}_{K}.)^T$ as in Lemma \ref{joint-influence}.
Let $\bm r := \Big(o_p(n^{-\frac12}), ..., o_p(n^{-\frac12})\Big)^T \in \mathbb{R}^{2K}$. 

By the Cramér-Wold theorem, $\sqrt{n}(\hat{\bm \Delta} - \bm \Delta^*) \Rightarrow N\big(\bm 0, \bm \Sigma_{\bar{\bm \phi}^*})$ if and only if $\forall \bm a \in \mathbb{R}^{2K}$, $\bm a^T\sqrt{n}(\hat{\bm \Delta} - \bm \Delta^*) \Rightarrow N\big(0, \bm a^T \bm \Sigma_{\bar{\bm \phi}^*} \bm a\big)$, i.e., we can prove convergence in distribution by proving that all linear combinations of the vector converge in distribution to a univariate Gaussian distribution. By the definition of asymptotic linearity, $\sqrt{n}(\bar{\bm \phi}^* + \bm r) = \sqrt{n}(\hat{\bm \Delta} - \bm \Delta^*)$. We want to show that all linear combinations of $\sqrt{n} (\bar{\bm \phi}^* + \bm r)$ converge in distribution, because that implies that $\sqrt{n}(\hat{\bm \Delta} - \bm \Delta^*)$ converges in distribution.

Take any $\bm a \in \mathbb{R}^{k}$. Then $\bm a^T (\bar{\bm \phi}^* + \bm r) = \bm a^T \bar{\bm \phi}^* + \bm a^T \bm r$. From Lemma \ref{joint-influence}, $\sqrt{n}\bar{\bm \phi}^* \Rightarrow N(\bm 0, \bm \Sigma_{\bar{\bm \phi}^*})$, and by the properties of the multivariate Gaussian distribution, $\bm a^T \sqrt{n}(\bar{\bm \phi}^* - \bm \phi) \Rightarrow N(0, \bm a^T \bm \Sigma_{\bar{\bm \phi}^*} \bm a)$.
$\bm a^T \bm r = \sum_{i=1}^{2K} a_i o_p(1)$. Since each element of the summation is $o_p(1)$, and since the summation has a finite number of elements, then $\bm a^T \bm r = o_p(1)$, i.e., $\bm a^T \bm r \pa 0$. By Slutsky's theorem, $\bm a^T \bar{\bm \phi}^* + \bm a^T \bm r \Rightarrow N(0, \bm a^T \bm \Sigma_{\bar{\bm \phi}^*} \bm a)$, which, since this holds $\forall \bm a$, by the Cramér-Wold theorem, $\sqrt{n} (\hat{\bm \Delta} - \bm \Delta^*) \Rightarrow N\Big(\bm 0, \bm \Sigma_{\bar{\bm \phi}^*} \Big)$.

\end{proof}

\begin{proof}[Proof of Corollary \ref{joint-statistics}]
Define $\bm M := \diag{\sqrt{t^*_1}, ..., \sqrt{t^*_k}}$ and $\bm B := \diag{\tilde{\sigma}^{-1}, \sigma^{-1}}$, and $\bm D := \bm M \otimes \bm B$. Recognize that $\bm D$ is a symmetric matrix, and that
\begin{align*}
	\bm D = \diag{\tilde{\sigma}^{-1} \sqrt{t^*_1}, \sigma^{-1} \sqrt{t^*_1}, ..., \tilde{\sigma}^{-1} \sqrt{t^*_k}, \sigma^{-1} \sqrt{t^*_k}}.
\end{align*}
Let $\hat{\bm D}$ be the same as $\bm D$ but with $k$th diagonal entry as $\hat{\sigma}_k^{-1} \sqrt{t^*_k}$ and the $(k+1)$th diagonal entry as $(\hat{\sigma}^C_k)^{-1} \sqrt{t^*_1}$. That is, $\hat{\bm D}$ is $\bm D$ where the standard error estimates have replaced the asymptotic standard error. By \citet{wangAnalysisCovarianceRandomized2019}, $\hat{\bm D} \to \bm  D$ in probability since $\hat{\sigma}_k$ and $\hat{\sigma}^C_k$ are consistent estimators of their population counterparts.
Note that the asymptotic variance of $\sqrt{n} \hat{\Delta}_{k}$ is $\tilde{\sigma}^2 t^{*-1}_{k}$, and the asymptotic variance of $\sqrt{n} \hat{\Delta}^{C}_{k}$ is $\sigma^2 t^{*-1}_k$, so multiplying by $\bm D$ will standardize the random variable $\hat{\bm \Delta}$. In particular, for an ANOVA estimate at stage $k$, $Z^*_k = \sqrt{n_k} (\hat{\Delta}_k - \Delta^*) / \hat{\sigma}_k$, and for an ANCOVA estimate at stage $k$, $Z^{*C}_k = \sqrt{n_k} (\hat{\Delta}_k - \Delta^*) / \hat{\sigma}^C_k$. Since $\sqrt{t^*_k} = \sqrt{n^*_k}/\sqrt{n}$, multiplying $\sqrt{n}(\hat{\bm \Delta} - \bm \Delta^*)$ by $\hat{\bm D}$ will get us a vector of suitably standardized test statistics $\bm Z^*$.
% , i.e., $\bm Z^* = \bm D \sqrt{n} (\hat{\bm \Delta} - \bm \Delta^*)$

To find the distribution of $\bm Z^*$, we have by Slutsky's theorem with the multivariate normal
\begin{align*}
    \bm Z^* = \hat{\bm D} \sqrt{n} (\hat{\bm \Delta} - \bm \Delta^*) &\Rightarrow N(\bm 0, \bm D \bm \Sigma_{\hat{\bm \Delta}} \bm D^T).
\end{align*}
$\bm Z^*$ is a vector of suitably standardized test statistics under the null hypothesis that $\Delta^* = 0$. We now derive the form of the variance of $\bm Z^*$, i.e., $\bm \Sigma_{\bm Z^*} := \bm D \bm \Sigma_{\hat{\bm \Delta}} \bm D^T$.

Recall that $\bm \Sigma_{\hat{\bm \Delta}} = \bm T \otimes \Var[(\phi, \phi^C)^T]$. Define $\bm V := \Var[(\phi, \phi^C)^T]$. Then, using properties of the Kronecker product (see Theorem 3 of \citet{zhangKroneckerProductsTheir2013}),
\begin{align*}
    \bm D \bm \Sigma_{\hat{\bm \Delta}} \bm D^T &= (\bm M \otimes \bm B) (\bm T \otimes \bm V) (\bm M \otimes \bm B) \\
    &= ((\bm M \bm T) \otimes (\bm B \bm V)) (\bm M \otimes \bm B) \\
    &= (\bm M \bm T \bm M) \otimes (\bm B \bm V \bm B).
\end{align*}
We now compute $(\bm M \bm T \bm M)$ and $(\bm B \bm V \bm B)$. We have
\begin{align*}
    \bm M \bm T \bm M &= \diag{\sqrt{t^*_1}, ..., \sqrt{t^*_K}} \begin{pmatrix}
        t^{*,-1}_1 & t^{*,-1}_2 & \dots & t^{*,-1}_K \\
        t^{*,-1}_2 & t^{*,-1}_2 & \dots & t^{*,-1}_K \\
        \dots & \dots & \dots & \dots \\
        t^{*,-1}_K & t^{*,-1}_{K} & \dots & t^{*,-1}_{K}
    \end{pmatrix} \diag{\sqrt{t^*_1}, \dots, \sqrt{t^*_K}} \\
    &= \begin{pmatrix}
        1 & \sqrt{t^*_1}/\sqrt{t^*_2} & \sqrt{t^*_1}/\sqrt{t^*_3} & \dots & \sqrt{t^*_1}/\sqrt{t^*_K} \\
        \sqrt{t^*_1}/\sqrt{t^*_2} & 1 & \sqrt{t^*_2}/\sqrt{t^*_3} & \dots & \dots \\
        \sqrt{t^*_1}/\sqrt{t^*_3} & \sqrt{t^*_2}/\sqrt{t^*_3} & 1 & \dots & \dots \\
        \dots & \dots & \dots & \dots & \dots \\
        \sqrt{t^*_1}/\sqrt{t^*_K} & \dots & \dots & 1
    \end{pmatrix} \\
    \bm B \bm V \bm B &= \begin{pmatrix}
    \tilde{\sigma}^{-1} & 0 \\
    0 & \sigma^{-1}
    \end{pmatrix} \begin{pmatrix}
    \tilde{\sigma}^2 & \sigma^2 \\
    \sigma^2 & \sigma^2
    \end{pmatrix} \begin{pmatrix}
    \tilde{\sigma}^{-1} & 0 \\
    0 & \sigma^{-1}
    \end{pmatrix} \\
    &= \begin{pmatrix}
    1 & \sigma/\tilde{\sigma} \\
    \sigma/\tilde{\sigma} & 1
    \end{pmatrix} = \begin{pmatrix}
    1 & \rho \\
    \rho & 1
    \end{pmatrix}.
\end{align*}
Putting them together, we have $(\bm M \bm T \bm M) \otimes (\bm B \bm V \bm B) = \bm T' \otimes \begin{psmallmatrix}
    1 & \rho \\
    \rho & 1
    \end{psmallmatrix}$, i.e.,
\begin{align*}
    \bm Z^* \Rightarrow N\Big(0, \bm T' \otimes \begin{psmallmatrix}
    1 & \rho \\
    \rho & 1
    \end{psmallmatrix}\Big)
\end{align*}
where $\bm T'$ has entries $\bm T'_{k,k'} = \sqrt{\frac{t^*_{\min(k, k')}}{t^*_{\max(k, k')}}}$.
\end{proof}

\textit{Note: Under an alternative hypothesis, $\Delta^* \neq 0$. We can instead have $\bm \Delta^*$ be the treatment effect under the alternative hypothesis we are considering, for example, to construct adjusted point estimates and confidence intervals as in Section \ref{sec:estimation-inference}. Instead of comparing the test statistic to the mean-zero multivariate Gaussian above, we compare our  test statistic minus $\sqrt{n} \hat{\bm D} \bm \Delta^*$ to the mean-zero multivariate Gaussian above (or equivalently to a multivariate Gaussian with the same variance as above but with mean $\sqrt{n} \hat{\bm D} \bm \Delta^*$).}

\subsection{Interim Monitoring in the Hybrid Design}\label{app:monitoring}

For background details on interim monitoring in a group sequential design, see \citet{gernotGroupSequentialConfirmatory2016}. Briefly, monitoring boundaries for common forms like Pocock \citep{pocockGroupSequentialMethods1977} or O'Brien-Fleming \citep{obrienMultipleTestingProcedure1979} can be derived using the joint distribution of test statistics in the group sequential design. Specifically, the constant $c$ is chosen such that
$$ \lim_{n\to\infty} P(|Z_1^*| \geq c \text{ or } ... \text{ or } |Z_{K}^*| \geq c) = \alpha$$ for the Pocock design, and $$ \lim_{n\to\infty} P(|Z_1^*| \geq c \text{ or } ... \text{ or } |Z_{K}^*| \geq c/\sqrt{K}) = \alpha$$ for the O'Brien-Fleming design. In the standard group sequential (non-hybrid) design, under the null hypothesis, the joint distribution of $(Z_1^*, ..., Z_K^*)$ converges to a mean-zero multivariate Gaussian distribution with a covariance matrix determined by the information fractions. In Theorem \ref{joint-estimates} and Corollary \ref{joint-statistics}, we showed that we can similarly characterize the joint distribution of the test statistics under the null hypothesis for the hybrid design proposed in \ref{def:hybrid}. Therefore, to solve for the constants in Corollaries \ref{corr:unif-inflated} and \ref{corr:end-inflated}, we are able to use the same type of strategy, but where we have modified the joint distribution of the test statistics to reflect the hybrid design.

\begin{proof}[Proof of Corollaries \ref{corr:unif-inflated} and \ref{corr:end-inflated}]
    In Corollary \ref{joint-statistics}, we characterized the asymptotic joint distribution of the vector $\bm Z^* = (Z^*_1, Z^{*C}_1, ..., Z_K^*, Z^{*C}_K)^T$. Since sub-vectors of multivariate Gaussian vectors are also multivariate Gaussian, we can simply select the covariance elements corresponding to a sequence of un-adjusted and covariate-adjusted estimators in line with the hybrid design propsed in \ref{def:hybrid}. Specifically, the test statistic sequence for the hybrid design $(Z_1^*, ..., Z_{K-1}^*, Z_K^{*C})$ converges in distribution to the multivariate Gaussian distribution characterized by $\bm U^{\rho}$ defined in Corollary \ref{corr:unif-inflated}, under the null hypothesis. Therefore, by definition, since the constant $c$ satisfies equation \eqref{eq:obf-critical-value}, then $\{c, c / \sqrt{2}, ..., c / \sqrt{K}\}$ satisfies Definition \ref{def:critical} of asymptotically acceptable critical values. The same proof by definition applies for Corollary \ref{corr:end-inflated}. In this case, we first select the constant $c$ based on the joint distribution of $\bm U^{1}$, which is what the sub-vector $(Z_1^*, ..., Z_{K-1}^*, Z_K^{*C})$ converges to under the null hypothesis with no covariate adjustment (i.e., $\rho = 1$, these are the standard OBF boundaries), and then modify only the last critical value to account for $\rho$ using $\bm U^{\rho}$.
\end{proof}

\textit{A Note on Computation:} A trick that is done to compute the Gaussian probabilities above in non-hybrid designs is called \textit{recursive integration} \citep{gernotGroupSequentialConfirmatory2016, armitageRepeatedSignificanceTests1969}. Recursive integration allows for quick computation of the probabilities based on the fact that there are independent increments of sizes $n_1, ..., n_K$, similar to how we derived the joint distribution of $\sqrt{n} \bar{\bm \phi}^*$ by partitioning it into a vector with elements that sum only observations within stage $k$. In the hybrid ANOVA-ANCOVA case, we can no longer construct independent increments because of the correlation between ANOVA and ANCOVA within the same stage. Instead, we use software that allows us to compute probabilities of Gaussian polytopes numerically. Using standard statistical software, this integration is feasible as the vast majority of trials have several stages, rather than tens of stages.

\textit{Alpha-Spending}: Alpha-spending can be incorporated into our approach \citep{lanDiscreteSequentialBoundaries1983}. There are several scenarios where one might want to employ alpha-spending related to our paper. The first is when the information fractions are not known ahead of time, and where the hybrid design in Procedure \ref{def:hybrid} is used. This is a standard application of alpha-spending. Rather than solving for the critical values as in Corollaries \ref{corr:unif-inflated} and \ref{corr:end-inflated}, one would specify an alpha-spending function, and sequentially solve for the critical values as the trial progresses.

Specifically, the critical values are calculated on the fly based on a pre-specified amount of type I error (alpha) to be spent at each stage. Let $\alpha^*(t)$ be a strictly increasing function with $\alpha^*(0) = 0$ and $\alpha^*(1) = \alpha$. Then, for the hybrid design in Procedure \ref{def:hybrid}, we would solve for the following critical values $\{u_1, ..., u_K\}$ sequentially, at each stage based on the observed information fractions that define the joint distribution of the test statistics in Corollary \ref{joint-statistics}:
\begin{align*}
         {P}(|Z_1^*| \geq u_1) &= \alpha^*(t^*_1) \\
         {P}(|Z_2^{*}| \geq u_2, |Z_{1}^*| < u_{1}) &= \alpha^*(t^*_2) - \alpha^*(t^*_{1}) \\
         ... \\
 	P(|Z_K^{*C}| \geq u_K, |Z_{K-1}^*| < u_{K-1}, ..., |Z_{2}^*| < u_{2}, |Z_1^*| < u_1) &= \alpha^*(t^*_K) - \alpha^*(t^*_{K-1}).
 \end{align*}
If we used the O'Brien-Fleming alpha-spending function, we would end up with the boundary shape of Corollary \ref{corr:end-inflated}, since all of the inflation would occur at the last stage. Therefore, we could use this approach with known, or estimated, $\rho$.

A second application of alpha-spending would be to choose a different alpha-spending function that spends relatively more type I error in earlier stages. This may be desirable if one wanted boundaries like that in Corollary \ref{corr:unif-inflated}, but did not know $\rho$ initially. A third application of alpha-spending is to extend our methods to an arbitrary sequence -- possibly unknown ahead of time -- of un-adjusted and covariate-adjusted estimators throughout the stages of a group sequential trial, and where $\rho$ could be re-estimated throughout the course of the trial.

\subsection{Adjusted Estimation and Inference in the Hybrid Design}\label{app:orderings}

In a standard group sequential design, calculating adjusted p-values, confidence intervals, and point estimates comes down to being able to define and calculate $P((Z^*_{k'}, k') \succeq (z^*_k, k))$. That is, when is the observed test statistic $Z^*_{k'}$ and stopping stage $k$ ``more extreme'' than some arbitrary values $(z^*_k, k)$? Calculating $P((Z^*_{k'}, k') \succeq (z^*_k, k))$ requires three main things: a joint distribution of test statistics, critical values for each stage that define the stopping boundaries, and a sample space ordering that defines ``extreme''. Our contribution is to modify these three elements for the hybrid design, which we describe below. We provide more details about sample space orderings in Section \ref{sec:sample-space-orderings}. Finally, we review how computing $P((Z^*_{k'}, k') \succeq (z^*_k, k))$ can be used for adjusted estimation and inference that takes into account the group sequential design.

\subsubsection{Modifications for Estimation and Inference in the Hybrid Design}

\begin{enumerate}
    \item The joint distribution of test statistics across stages. In particular, we need to know the covariance matrix between the test statistics across stages and when performing ANOVA and ANCOVA at each stage. In Corollary \ref{joint-statistics}, we derived this covariance matrix. In practice, we would select the subset of the covariance matrix that represents the tests that are needed for a particular ordering, e.g., $(Z_1^*, Z_2^*, Z_3^*, Z_4^{*C})$.
    \item Critical values $(l_k, u_k)$ at each stage $k$. The pairing of these critical values with the joint distribution of test statistics in Corollary \ref{joint-statistics} allows us to characterize the distribution of observed test statistics. The distribution of the observed test statistics differs from the joint distribution in Corollary \ref{joint-statistics} in that only some values of the test statistics will be observed at particular stages, based on whether or not the null hypothesis was rejected at that stage. For example, if $(l_1, u_1) = (-4, 4)$, then the probability of observing $(z_1^* = 3, k=1)$ is exactly 0 because the null hypothesis cannot be rejected with a test statistic of 3 in the first stage. These critical values should be chosen so that they control type I error in the hybrid design, discussed in Section \ref{app:monitoring}.
    \item An sample space ordering for the two-dimensional vector $(z_k^*, k)$. The orderings tell us whether $(z_{k'}^*, k') \succeq (z_k^*, k)$. It is then used in conjunction with (1) and (2) to perform computations with the observed test statistic, i.e., $P((Z^*_{k'}, k') \succeq (z^*_k, k))$. We provide more information about sample space orderings and how we modify them for the hybrid design in Section \ref{sec:sample-space-orderings}.
\end{enumerate}

\subsubsection{Sample Space Orderings and their Modifications for the Hybrid Design}\label{sec:sample-space-orderings}

We briefly describe four commonly used sample space orderings and discuss two in detail: stage-wise ordering and sample mean ordering. Our contribution is to modify the computations for what is considered ``as or more extreme'' in the context of the hybrid ANOVA-ANCOVA analysis, which is then used to construct p-values, confidence intervals, and point estimates. 

Four sample space orderings are used in practice. Consider two test statistics $(z_k^*, k)$ and $(z_{k'}^*, k')$, and we want to say whether $(z_k^*, k) \succeq (z_{k'}^*, k')$. Additional details of these common sample space orderings can be found elsewhere \citep{gernotGroupSequentialConfirmatory2016}.
\begin{itemize}
    \item Stage-wise ordering is based on the value of the $z_k$'s alone if $k=k'$. $(z^*_{k'}, k') \succeq (z^*_k, k)$ if $k' < k$ and $z^*_{k'} \geq u_{k'}$ (crossing the upper boundary at an earlier stage), and $(z^*_{k'}, k') \preceq (z^*_k, k)$ if $k' < k$ and $z^*_{k'} \leq u_{k'}$ (crossing the lower boundary at an earlier stage).
    \item Likelihood ratio ordering has $(z^*_{k'}, k') \succeq (z^*_k, k)$ if $z^*_{k'} > z^*_k$.
    \item Sample mean ordering ignores stage and sample size altogether and just compares the values of the standardized sample means, rather than the test statistics. If each stage has an equal sample size, this reduces to considering $(z^*_{k'}, k') \succeq (z^*_k, k)$ if $\frac{z^*_{k'}}{\sqrt{k'}} > \frac{z^*_k}{\sqrt{k}}$.
    \item Score test ordering orders test statistics based on the score statistic. If each stage has an equal sample size, this reduces to considering $(z^*_{k'}, k') \succeq (z^*_k, k)$ if $z^*_{k'} \sqrt{k'} > z^*_k \sqrt{k}$.
\end{itemize}

For the hybrid ANOVA-ANCOVA design that we consider, the ordering is based on the ANCOVA test at the last stage, whether that be from stopping early or from continuing through the last stage. The key difference between performing standard ANOVA or ANCOVA analyses at each stage and a hybrid ANOVA-ANCOVA analysis is that since the ordering is based off of ANCOVA at the last stage, in the hybrid setting, there may be two tests performed at the same stage (i.e., an ANOVA that crosses an interim boundary followed by an ANCOVA).

Now we detail the computations of $ P[(Z_{k'}^*, k') \succeq (z^*_k, k)]$ for two of these orderings and describe the modifications that we need to make in order to account for hybrid ANOVA-ANCOVA analyses. In both settings, we focus on the setting where the trial is monitored with an ANOVA test statistic, $Z^*_k$, and estimation and inference is desired based on a final ANCOVA test $Z^{*C}_k$. Let $C_j$ be the event of not rejecting through stage $j$ using an ANOVA statistic, and $R_j$ be the event of rejecting at stage $j$ using an ANOVA statistic. We can obtain the probabilities below using the covariance of $\bm Z^*$ from Corollary \ref{joint-statistics}, which is a function of both the information fraction (and sometimes the joint distribution between ANOVA and ANCOVA). 

\textit{Stage-Wise Ordering}

When using the stage-wise ordering, rejecting at a later stage is considered a \textit{smaller} value, so we don't consider it in the upper tail probabilities. We note that for the non-hybrid and hybrid designs, we only need the distribution of $Z_k^*$ and $Z_k^{*C}$ values at stages $1, ..., k$. We do not need the distribution for stages $k+1, ..., K$. Furthermore, the hybrid designs require the joint distribution of $Z_k^*$ and $Z_k^{*C}$.

In the non-hybrid design, we perform ANOVA tests in the interim and we wish to perform estimation and inference using ANOVA. The following shows how to compute $P[(Z_{k'}^*, k') \succeq (z^*_k, k)]$ in this setting. It is written for ANOVA, but also covers ANCOVA just by replacing $Z^*$ with $Z^{*C}$ since the covariance is the same as long as we're not switching between ANOVA and ANCOVA.
\begin{align*}
    P[(Z_{k'}^*, k') \succeq (z^*_k, k)] &= \sum_{j=1}^{k-1} \Big\{P[Z^*_j \geq u_j | C_{j-1}, R_j] P[C_{j-1}, R_j] \Big\} + \\
        & \quad \quad \quad P[Z_{k}^* \geq z_k^* | C_{k-1}] P[C_{k-1}] \\
        &= \sum_{j=1}^{k-1} \Big\{P[Z^*_j \geq u_j, (Z^*_j \geq u_j \text{ or } Z^*_j \leq \ell_j), \ell_{j-1} < Z^*_{j-1} < u_{j-1}, ..., \ell_{1} < Z^*_{1} < u_{1}] \Big\} + \\
        & \quad \quad \quad P[Z_{k}^* \geq z_k^*,  \ell_{k-1} < Z^*_{k-1} < u_{k-1}, ..., \ell_{1} < Z^*_{1} < u_{1}] \\
        &= \sum_{j=1}^{k-1} \Big\{P[Z^*_j \geq u_j, \ell_{j-1} < Z^*_{j-1} < u_{j-1}, ..., \ell_{1} < Z^*_{1} < u_{1}] \Big\} + \\
        & \quad \quad \quad P[Z_{k}^* \geq z_k^*,  \ell_{k-1} < Z^*_{k-1} < u_{k-1}, ..., \ell_{1} < Z^*_{1} < u_{1}].
\end{align*}

In the hybrid design, there are two options. The first scenario is that the trial is stopped early after crossing a boundary using an ANOVA statistic, and an ANCOVA analysis is performed at that same stage. The second scenario is that the trial continues until the last stage (i.e., no boundaries are crossed during interim monitoring), and only an ANCOVA analysis is performed. The computation for $P[(Z_{k'}^*, k') \succeq (z^*_k, k)]$ is different between these cases:
\begin{itemize}
\item Early Stopping
    \begin{align*}
        P[(Z_{k'}^{*C}, k') \succeq (z^{*C}_k, k)] &= \sum_{j=1}^{k-1} \Big\{P[Z^*_j \geq u_j | C_{j-1}, R_j] P[C_{j-1}, R_j] \Big\} + \\
            & \quad \quad \quad P[Z_{k}^{*C} \geq z_k^{*C} | R_{k}, C_{k-1}] P[R_{k}, C_{k-1}] \\
            &= \sum_{j=1}^{k-1} \Big\{P[Z^*_j \geq u_j, \ell_{j-1} < Z^*_{j-1} < u_{j-1}, ..., \ell_{1} < Z^*_{1} < u_{1}] \Big\} + \\
            & \quad \quad \quad P[Z_{k}^{*C} \geq z_k^{*C}, (Z^*_{k} \geq u_{k} \text{ or } Z^*_{k} \leq \ell_{k}), ..., \ell_{1} < Z^*_{1} < u_{1}].
    \end{align*}
\item No Early Stopping
    \begin{align*}
        P[(Z_{k'}^{*C}, k') \succeq (z^{*C}_k, k)] &= \sum_{j=1}^{k-1} \Big\{P[Z^*_j \geq u_j | C_{j-1}, R_j] P[C_{j-1}, R_j] \Big\} + \\
            & \quad \quad \quad P[Z_{k}^{*C} \geq z_k^{*C} | C_{k-1}] P[C_{k-1}] \\
            &= \sum_{j=1}^{k-1} \Big\{P[Z^*_j \geq u_j, \ell_{j-1} < Z^*_{j-1} < u_{j-1}, ..., \ell_{1} < Z^*_{1} < u_{1}] \Big\} + \\
            & \quad \quad \quad P[Z_{k}^{*C} \geq z_k^{*C},  \ell_{k-1} < Z^*_{k-1} < u_{k-1}, ..., \ell_{1} < Z^*_{1} < u_{1}].
    \end{align*}
\end{itemize}
    
\textit{Sample Mean Ordering}

Let $X_k^*$ be the cumulative sample mean (i.e., the parameter estimate), where $X_k^* = Z_k^* / \sqrt{n^*_k}$. Let $X^{*C}_k$ be the sample mean using ANCOVA. Let $(x^*_k, k)$ be the observed bivariate test statistic for stopping stage and sample mean. Sample mean ordering is not directly a function of the stopping stage, so we ignore the observed stopping stage in the calculations, other than to note that we are at the end of the trial. Note that rather than using $\bm \Sigma_{\bm Z^*}$ as the covariance, we use the covariance matrix for the sample means in Theorem \ref{joint-estimates}, since we can translate the rejection region to the sample mean scale rather than the likelihood ratio scale.

This is written for ANOVA, but also covers ANCOVA just by replacing $Z^*$ with $Z^{*C}$ and $X^*$ with $X^{*C}$ since the covariance is the same as long as we're not switching between ANOVA and ANCOVA.
        \begin{align*}
            P[X^*_{k'} \succeq x_k^*] &= \sum_{j=1}^{K-1} \Big\{ P[X^*_j \geq x^*_j | R_j, C_{j-1}] P[R_j, C_{j-1}] \Big\} + \\
            & \quad \quad \quad P[X_{K}^{*} \geq x_k^{*} | C_{K-1}] P[C_{K-1}] \\
            &= \sum_{j=1}^{K-1} \Big\{P[X^*_j \geq x^*_k, (Z^*_{j} \geq u_{j} \text{ or } Z^*_{j} \leq \ell_{j}), \ell_{j-1} < Z^*_{j-1} < u_{j-1}, ..., \ell_{1} < Z^*_{1} < u_{1}] \Big\} + \\
                & \quad \quad \quad P[X_{K}^* \geq x_k^*, \ell_{K-1} < Z^*_{K-1} < u_{K-1}, ..., \ell_{1} < Z^*_{1} < u_{1}].
        \end{align*}

Since the sample mean ordering is not directly a function of the stopping stage, the computation for scenarios where there is early stopping versus not can be grouped together.
\begin{align*}
    P[X^{*C}_{k'} \succeq x_k^{*C}] &= \sum_{j=1}^{K-1} \Big\{ P[X^{*C}_j \geq x^{*C}_k | R_j, C_{j-1}] P[R_j, C_{j-1}] \Big\} + \\
    & \quad \quad \quad P[X_{K}^{*C} \geq x_k^{*C} | C_{K-1}] P[C_{K-1}] \\
    &= \sum_{j=1}^{K-1} \Big\{P[X^{*C}_j \geq x^{*C}_k, (Z^*_{j} \geq u_{j} \text{ or } Z^*_{j} \leq \ell_{j}), \ell_{j-1} < Z^*_{j-1} < u_{j-1}, ..., \ell_{1} < Z^*_{1} < u_{1}] \Big\} + \\
        & \quad \quad \quad P[X_{K}^{*C} \geq x_k^{*C}, \ell_{K-1} < Z^*_{K-1} < u_{K-1}, ..., \ell_{1} < Z^*_{1} < u_{1}].
\end{align*}

\subsubsection{Review of Estimation and Inference in Group Sequential Designs}\label{app:construction}

Up until now, we have modified the sample space ordering and related computation for $P((Z^*_{k'}, k') \succeq (z^*_k, k))$. To implement these probabilities, we use the results from Theorem \ref{joint-estimates} and Corollary \ref{joint-statistics}. Now, we can calculate $p$-values, the probability of observing $(z^*_k, k)$ or something \textit{more extreme} under the null hypothesis $P_0$ using a standard group sequential approach \citep{gernotGroupSequentialConfirmatory2016}:
\begin{align}\label{simple-conf}
\begin{split}
    P_{0}((Z^*_{k'}, k') \succeq (z^*_k, k)).
\end{split}
\end{align}
A two-sided p-value can be computed similarly, using both the above and $P_{0}((Z^*_{k'}, k') \preceq (z^*_k, k))$. We can use these probabilities to compute a $(1-\alpha)*100$\% confidence interval, we find $\Delta_l$ and $\Delta_u$ effect sizes for $P$ such that
\begin{align*}
    P_{\Delta_l}((Z_{k'}^*, k') \succeq (z_k^*, k)) =
    P_{\Delta_u}((Z_{k'}^*, k') \preceq (z_k^*, k)) = \alpha/2.
\end{align*}

Then $(\Delta_l, \Delta_u)$ is a $(1-\alpha)*100$\% confidence interval \citep{gernotGroupSequentialConfirmatory2016}. In a similar fashion, we can derive median unbiased point estimates. We define $\tilde{\Delta}$ as the median unbiased point estimate \citep{gernotGroupSequentialConfirmatory2016}, which is the value that satisfies
\begin{align*}
    P_{\tilde{\Delta}}((Z_{k'}^*, k') \succeq (z_k^*, k)) =
    P_{\tilde{\Delta}}((Z_{k'}^*, k') \succeq (z_k^*, k)) = 0.5.
\end{align*}

In all of these cases, we use the covariance matrix of $\bm Z^*$ described in Corollary \ref{joint-statistics} to compute these probabilities. When working with $P_0$, we use the distribution exactly as described in Corollary \ref{joint-statistics}; when working with $P_{\Delta}$ for some non-zero effect size, we suitably center and scale the test statistics first based on the alternative hypothesis of effect size $\Delta$, then use Corollary \ref{joint-statistics}.

\newpage
\section{Simulation Methods and Results}\label{app:simulation}

\subsection{Simulation Methods}\label{app:sim-meth}

We make the following assumptions for our simulation: the overall variance, $\tilde{\sigma}^2 = 1$, the variance of each covariate $\Var(X_i) = 1$ all covariates are independent of one another, and all of the covariates have the same coefficient, i.e., $\bm \gamma = \gamma \bm 1$. For simplicity, we set $\E(X_i) = 0$ for each covariate. With that, we input the following parameters: $p$ the number of covariates and $\rho$. Given these two parameters, we solve for $\gamma$ (the coefficients) and $\sigma^2$ (the observation variance). Since $\rho^2 = \sigma^2 / \tilde{\sigma}^2$, the observation variance $\sigma^2 = \rho^2$. Furthermore, we want $\Var[\bm X^T \bm \gamma] = 1 - \rho^2$. Using our assumptions about the covariates, we have
\begin{align*}
	\Var[\bm X^T \bm \gamma] &= \sum_{j=1}^{p} \gamma^2 \Var[X_j] \\
	&= p\gamma^2 := 1-\rho^2 \\
	&\implies \gamma = \sqrt{\frac{1-\rho^2}{p}}.
\end{align*}

\newpage
\subsection{Supplemental Simulation Results}\label{app:supp-sim}

We include additional results for simulations with different parameters than were included in the main text. Below is a list of table and figure contents:
\begin{itemize}
	\item \textbf{Figure A1}: Comparison of type I error for strategy (B)(ii) with misspecified values of $\rho$ during trial design.
	\item \textbf{Table \ref{tab:bias-cover-obf-3}}: Median/mean scaled bias and coverage under OBF boundaries, 3-stage trial, one covariate. This is the same as Table \ref{tab:bias-cover}, but with OBF bounds instead of Pocock.
	\item \textbf{Table \ref{tab:bias-cover-pocock-2}}: Median/mean scaled bias and coverage with Pocock boundaries, 2-stage trial, one covariate. This is the same as Table \ref{tab:bias-cover}, but with 2 equally-sized stages instead of 3.
	\item \textbf{Table \ref{tab:bias-cover-pocock-cov2}}: Median/mean scaled bias and coverage with Pocock boundaries, 3-stage trial, two covariates. This is the same as Table \ref{tab:bias-cover}, but with two covariates instead of one to assess the impact of estimating multiple coefficients in finite samples. The $\Var[\bm X^T \bm \gamma]$ is still fixed to be $1-\rho^2$, but the coefficients $\bm \gamma$ are equal to the $\gamma$ in Section \ref{app:sim-meth}.
\end{itemize}

\begin{figure}[htbp!]
	\caption{Type I error under scenario (B)(ii) with misspecified $\rho$ at the design stage across 25,000 simulated trials. Stars indicate where the anticipated reduction and the true reduction in variance from using ANCOVA match up (correct specification), which represents the orange line in Figure 4. The dark purple line at the top of each subplot represents anticipating that there is no reduction in variance from using ANCOVA, equivalent to not modifying the boundaries, which represents the green line in Figure 4.}\label{fig:type1error-app}
    \includegraphics[width=\textwidth]{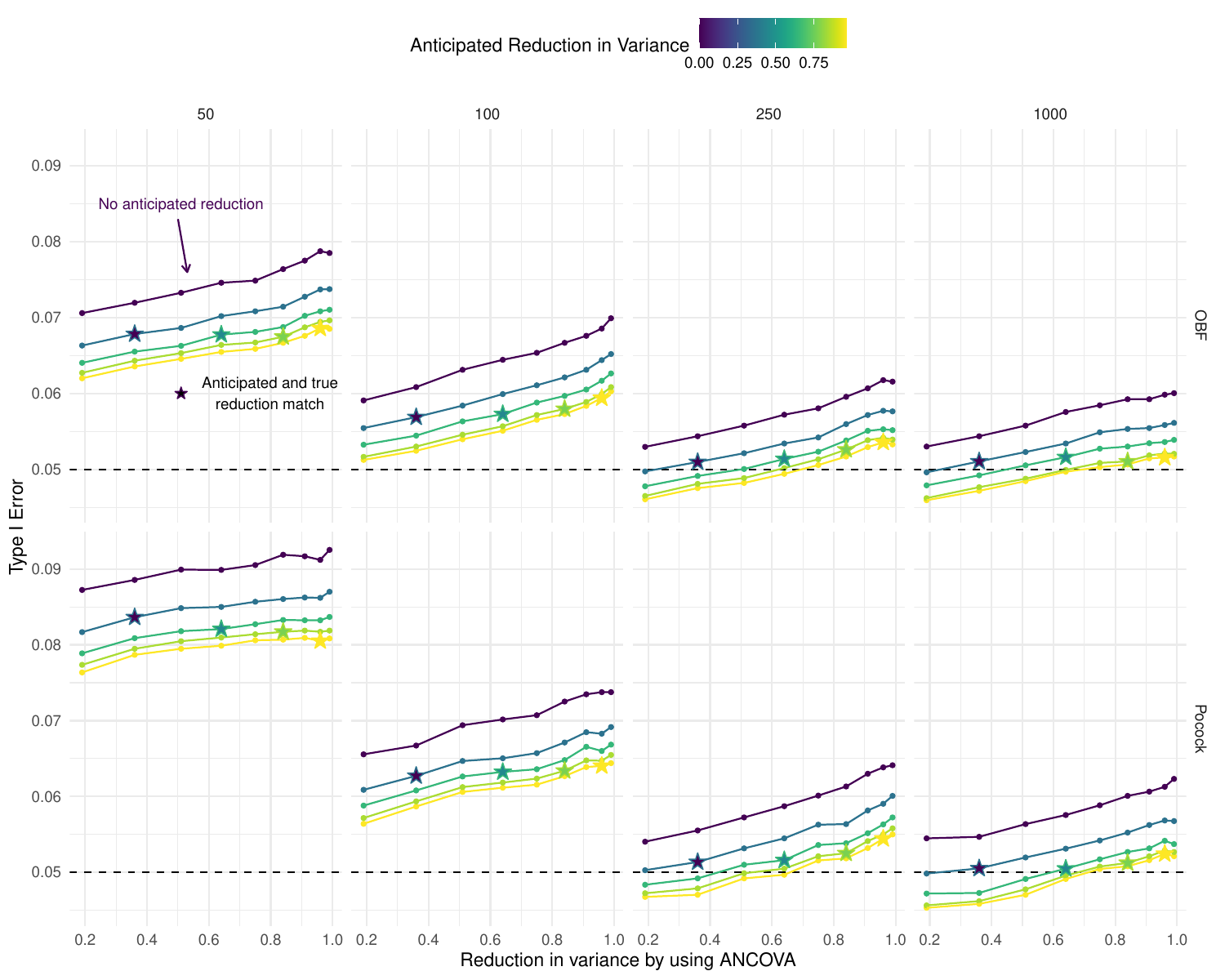}
\end{figure}

% latex table generated in R 4.1.2 by xtable 1.8-4 package
% Tue May 24 10:26:23 2022
\begin{table}[htbp!]
\caption{Bias ($\times 100$) and coverage results based on 10,000 simulated trials for each row, end-inflated OBF boundaries for 3-stage trial. Bias is shown as median bias and mean bias in parentheses. $\Delta$ is the true effect size, $\rho$ is the true ratio of standard error between the ANCOVA and ANOVA estimators, and $n$ is the total sample size for the trial.}\label{tab:bias-cover-obf-3}
\begin{tabular}{ccc|ccc|ccc}
  \hline

  \hline
& & & \multicolumn{3}{c|}{Scaled Bias: Median (Mean)} & \multicolumn{3}{c}{Coverage} \\
\hline
$\Delta$ & $\rho$ & $n$ & Simple & GS & GS + Adjust & Simple & GS & GS + Adjust \\ 
  \hline
0.00 & 0.25 &  50 & -0.03 (-0.02) & -0.11 (-0.18) & -0.03 (-0.10) & 0.93 & 0.92 & 0.93 \\ 
  0.00 & 0.25 & 100 & 0.01 (-0.00) & -0.04 (-0.11) & 0.02 (-0.01) & 0.94 & 0.93 & 0.94 \\ 
  0.00 & 0.25 & 250 & 0.02 (0.01) & -0.01 (-0.07) & 0.02 (-0.00) & 0.95 & 0.94 & 0.94 \\ 
  0.00 & 0.25 & 1000 & -0.00 (-0.00) & -0.01 (-0.03) & 0.00 (-0.01) & 0.95 & 0.94 & 0.95 \\ 
   \hline 
0.00 & 0.50 &  50 & -0.04 (-0.02) & -0.16 (-0.29) & -0.04 (-0.06) & 0.93 & 0.92 & 0.93 \\ 
  0.00 & 0.50 & 100 & 0.03 (0.00) & -0.06 (-0.18) & 0.03 (-0.01) & 0.94 & 0.93 & 0.94 \\ 
  0.00 & 0.50 & 250 & 0.04 (0.01) & -0.01 (-0.10) & 0.04 (0.02) & 0.95 & 0.94 & 0.95 \\ 
  0.00 & 0.50 & 1000 & -0.00 (-0.01) & -0.03 (-0.06) & -0.01 (-0.02) & 0.95 & 0.94 & 0.95 \\ 
   \hline 
0.10 & 0.25 &  50 & -0.01 (0.04) & -0.93 (-1.54) & 0.00 (0.92) & 0.93 & 0.95 & 0.94 \\ 
  0.10 & 0.25 & 100 & 0.03 (0.03) & -2.60 (-2.64) & 0.06 (0.98) & 0.94 & 0.96 & 0.94 \\ 
  0.10 & 0.25 & 250 & 0.03 (0.04) & -5.14 (-4.65) & 0.03 (0.87) & 0.95 & 0.11 & 0.95 \\ 
  0.10 & 0.25 & 1000 & 0.01 (0.04) & -5.07 (-5.95) & -0.00 (0.28) & 0.95 & 0.03 & 0.95 \\ 
   \hline 
0.10 & 0.50 &  50 & 0.00 (0.19) & -0.30 (-0.81) & -0.01 (0.79) & 0.93 & 0.94 & 0.94 \\ 
  0.10 & 0.50 & 100 & 0.07 (0.12) & -0.24 (-1.03) & 0.07 (0.71) & 0.94 & 0.94 & 0.94 \\ 
  0.10 & 0.50 & 250 & 0.09 (0.15) & -1.06 (-1.54) & 0.06 (0.68) & 0.95 & 0.96 & 0.95 \\ 
  0.10 & 0.50 & 1000 & 0.07 (0.16) & -2.13 (-2.60) & -0.02 (0.30) & 0.95 & 0.52 & 0.95 \\ 
   \hline 
0.20 & 0.25 &  50 & 0.03 (0.11) & -9.10 (-8.21) & 0.10 (2.12) & 0.93 & 0.38 & 0.93 \\ 
  0.20 & 0.25 & 100 & 0.03 (0.09) & -12.19 (-10.57) & 0.10 (1.78) & 0.94 & 0.20 & 0.94 \\ 
  0.20 & 0.25 & 250 & 0.07 (0.09) & -10.68 (-11.94) & -0.21 (0.54) & 0.95 & 0.03 & 0.95 \\ 
  0.20 & 0.25 & 1000 & 0.03 (0.07) & -14.68 (-7.98) & -0.08 (0.21) & 0.95 & 0.43 & 0.95 \\ 
   \hline 
0.20 & 0.50 &  50 & 0.13 (0.42) & -1.79 (-2.74) & 0.11 (1.67) & 0.93 & 0.94 & 0.93 \\ 
  0.20 & 0.50 & 100 & 0.17 (0.35) & -4.98 (-4.01) & 0.15 (1.39) & 0.94 & 0.95 & 0.94 \\ 
  0.20 & 0.50 & 250 & 0.23 (0.36) & -4.55 (-5.26) & -0.16 (0.57) & 0.94 & 0.52 & 0.94 \\ 
  0.20 & 0.50 & 1000 & 0.14 (0.27) & -9.36 (-4.79) & -0.12 (0.28) & 0.95 & 0.43 & 0.95 \\ 
   \hline 
0.50 & 0.25 &  50 & 0.14 (0.32) & -29.23 (-28.71) & -0.17 (0.89) & 0.93 & 0.08 & 0.94 \\ 
  0.50 & 0.25 & 100 & 0.12 (0.25) & -33.36 (-26.01) & 0.21 (1.10) & 0.94 & 0.22 & 0.94 \\ 
  0.50 & 0.25 & 250 & 0.06 (0.11) & -0.78 (-8.80) & -0.05 (-1.20) & 0.95 & 0.73 & 0.95 \\ 
  0.50 & 0.25 & 1000 & -0.02 (-0.01) & -0.02 (-0.01) & -0.02 (-0.01) & 0.95 & 0.95 & 0.95 \\ 
   \hline 
0.50 & 0.50 &  50 & 0.58 (1.26) & -10.30 (-12.75) & -0.05 (1.33) & 0.92 & 0.61 & 0.93 \\ 
  0.50 & 0.50 & 100 & 0.42 (0.89) & -16.76 (-13.11) & 0.40 (1.29) & 0.94 & 0.50 & 0.94 \\ 
  0.50 & 0.50 & 250 & 0.23 (0.40) & -1.02 (-5.87) & -0.10 (-0.61) & 0.95 & 0.74 & 0.95 \\ 
  0.50 & 0.50 & 1000 & -0.04 (-0.02) & -0.04 (-0.02) & -0.04 (-0.02) & 0.95 & 0.95 & 0.95 \\ 
   \hline 
 \hline
\end{tabular}
\end{table}

% latex table generated in R 4.1.2 by xtable 1.8-4 package
% Tue May 24 10:26:23 2022
\begin{table}[htbp!]
\caption{Bias ($\times 100$) and coverage results based on 10,000 simulated trials for each row, end-inflated Pocock boundaries, 2-stage trial. Bias is shown as median bias and mean bias in parentheses. $\Delta$ is the true effect size, $\rho$ is the true ratio of standard error between the ANCOVA and ANOVA estimators, and $n$ is the total sample size for the trial.}\label{tab:bias-cover-pocock-2}
\begin{tabular}{ccc|ccc|ccc}
  \hline
  \hline
& & & \multicolumn{3}{c|}{Scaled Bias: Median (Mean)} & \multicolumn{3}{c}{Coverage} \\
\hline
$\Delta$ & $\rho$ & $n$ & Simple & GS & GS + Adjust & Simple & GS & GS + Adjust \\ 
  \hline
0.00 & 0.25 &  50 & 0.02 (0.03) & 0.02 (0.03) & 0.03 (0.13) & 0.94 & 0.96 & 0.93 \\
  0.00 & 0.25 & 100 & 0.02 (0.01) & 0.02 (0.01) & 0.03 (0.07) & 0.94 & 0.97 & 0.94 \\
  0.00 & 0.25 & 250 & 0.02 (0.01) & 0.02 (0.01) & 0.02 (-0.05) & 0.95 & 0.97 & 0.95 \\
  0.00 & 0.25 & 1000 & 0.01 (0.01) & 0.01 (0.01) & 0.01 (0.01) & 0.95 & 0.97 & 0.95 \\
   \hline
0.00 & 0.50 &  50 & 0.04 (0.07) & 0.04 (0.07) & 0.05 (0.19) & 0.93 & 0.96 & 0.93 \\
  0.00 & 0.50 & 100 & 0.05 (0.01) & 0.05 (0.01) & 0.05 (-0.01) & 0.94 & 0.96 & 0.94 \\
  0.00 & 0.50 & 250 & 0.04 (0.03) & 0.04 (0.03) & 0.04 (-0.03) & 0.94 & 0.97 & 0.94 \\
  0.00 & 0.50 & 1000 & 0.02 (0.02) & 0.02 (0.02) & 0.02 (0.01) & 0.95 & 0.97 & 0.95 \\
   \hline
0.10 & 0.25 &  50 & 0.06 (0.12) & -1.12 (-1.36) & 0.09 (1.39) & 0.94 & 0.97 & 0.94 \\
  0.10 & 0.25 & 100 & 0.05 (0.07) & -2.70 (-2.52) & 0.04 (1.23) & 0.94 & 0.96 & 0.95 \\
  0.10 & 0.25 & 250 & 0.05 (0.09) & -5.13 (-4.25) & 0.04 (1.07) & 0.95 & 0.16 & 0.95 \\
  0.10 & 0.25 & 1000 & 0.03 (0.06) & -1.46 (-3.44) & 0.01 (0.07) & 0.95 & 0.50 & 0.95 \\
   \hline
0.10 & 0.50 &  50 & 0.15 (0.36) & -0.01 (-0.15) & 0.14 (1.21) & 0.94 & 0.96 & 0.94 \\
  0.10 & 0.50 & 100 & 0.15 (0.28) & -0.20 (-0.47) & 0.06 (1.01) & 0.94 & 0.96 & 0.94 \\
  0.10 & 0.50 & 250 & 0.16 (0.31) & -1.25 (-1.15) & 0.08 (0.90) & 0.94 & 0.96 & 0.95 \\
  0.10 & 0.50 & 1000 & 0.13 (0.24) & -2.19 (-1.97) & 0.04 (0.19) & 0.95 & 0.51 & 0.95 \\
   \hline
0.20 & 0.25 &  50 & 0.08 (0.19) & -9.09 (-7.63) & 0.11 (2.32) & 0.94 & 0.58 & 0.94 \\
  0.20 & 0.25 & 100 & 0.08 (0.16) & -12.17 (-9.11) & 0.12 (1.75) & 0.94 & 0.23 & 0.95 \\
  0.20 & 0.25 & 250 & 0.06 (0.13) & -2.99 (-6.84) & -0.06 (0.13) & 0.95 & 0.50 & 0.95 \\
  0.20 & 0.25 & 1000 & 0.03 (0.02) & 0.01 (-0.17) & 0.02 (-0.08) & 0.95 & 0.94 & 0.95 \\
   \hline
0.20 & 0.50 &  50 & 0.30 (0.67) & -2.06 (-1.84) & 0.22 (1.92) & 0.93 & 0.95 & 0.94 \\
  0.20 & 0.50 & 100 & 0.29 (0.59) & -5.13 (-3.38) & 0.13 (1.53) & 0.94 & 0.95 & 0.94 \\
  0.20 & 0.50 & 250 & 0.27 (0.48) & -4.68 (-3.92) & -0.09 (0.36) & 0.95 & 0.52 & 0.95 \\
  0.20 & 0.50 & 1000 & 0.05 (0.05) & 0.03 (-0.10) & 0.04 (-0.04) & 0.95 & 0.94 & 0.95 \\
   \hline
0.50 & 0.25 &  50 & 0.10 (0.21) & -3.37 (-13.93) & -0.48 (-0.89) & 0.93 & 0.59 & 0.94 \\
  0.50 & 0.25 & 100 & -0.01 (0.05) & -0.36 (-3.94) & -0.07 (-1.40) & 0.94 & 0.85 & 0.94 \\
  0.50 & 0.25 & 250 & 0.00 (0.01) & 0.00 (-0.00) & 0.00 (0.00) & 0.95 & 0.94 & 0.94 \\
  0.50 & 0.25 & 1000 & 0.02 (0.01) & 0.02 (0.01) & 0.02 (0.01) & 0.95 & 0.95 & 0.95 \\
   \hline
0.50 & 0.50 &  50 & 0.40 (0.88) & -4.76 (-8.60) & -0.58 (-0.11) & 0.93 & 0.60 & 0.94 \\
  0.50 & 0.50 & 100 & 0.03 (0.24) & -0.41 (-2.74) & -0.09 (-0.86) & 0.94 & 0.86 & 0.94 \\
  0.50 & 0.50 & 250 & 0.01 (0.03) & 0.00 (0.01) & 0.00 (0.02) & 0.95 & 0.94 & 0.94 \\
  0.50 & 0.50 & 1000 & 0.04 (0.03) & 0.04 (0.03) & 0.04 (0.03) & 0.95 & 0.95 & 0.95 \\
   \hline 
 \hline
\end{tabular}
\end{table}

% latex table generated in R 4.1.2 by xtable 1.8-4 package
% Tue May 24 10:26:23 2022
\begin{table}[htbp!]
\caption{Bias ($\times 100$) and coverage results based on 10,000 simulated trials for each row, end-inflated Pocock boundaries, 3-stage trial, 2 covariates. Bias is shown as median bias and mean bias in parentheses. $\Delta$ is the true effect size, $\rho$ is the true ratio of standard error between the ANCOVA and ANOVA estimators, and $n$ is the total sample size for the trial.}\label{tab:bias-cover-pocock-cov2}
\begin{tabular}{ccc|ccc|ccc}
  \hline
  \hline
& & & \multicolumn{3}{c|}{Scaled Bias: Median (Mean)} & \multicolumn{3}{c}{Coverage} \\
\hline
$\Delta$ & $\rho$ & $n$ & Simple & GS & GS + Adjust & Simple & GS & GS + Adjust \\ 
  \hline
0.00 & 0.25 &  50 & -0.04 (-0.01) & -0.14 (-0.18) & -0.07 (-0.29) & 0.93 & 0.94 & 0.91 \\ 
  0.00 & 0.25 & 100 & -0.00 (-0.01) & -0.05 (-0.12) & 0.01 (-0.02) & 0.94 & 0.96 & 0.94 \\ 
  0.00 & 0.25 & 250 & -0.02 (-0.01) & -0.05 (-0.08) & -0.02 (-0.03) & 0.94 & 0.96 & 0.94 \\ 
  0.00 & 0.25 & 1000 & -0.01 (0.00) & -0.02 (-0.03) & -0.01 (-0.01) & 0.95 & 0.96 & 0.95 \\ 
   \hline 
0.00 & 0.50 &  50 & -0.09 (-0.00) & -0.28 (-0.33) & -0.09 (-0.14) & 0.92 & 0.93 & 0.91 \\ 
  0.00 & 0.50 & 100 & -0.01 (-0.00) & -0.10 (-0.21) & 0.00 (-0.02) & 0.94 & 0.95 & 0.94 \\ 
  0.00 & 0.50 & 250 & -0.04 (-0.04) & -0.11 (-0.17) & -0.04 (-0.04) & 0.94 & 0.95 & 0.94 \\ 
  0.00 & 0.50 & 1000 & -0.02 (0.01) & -0.05 (-0.05) & -0.02 (0.01) & 0.94 & 0.96 & 0.95 \\ 
   \hline 
0.10 & 0.25 &  50 & 0.00 (0.12) & -1.63 (-1.92) & 0.02 (1.79) & 0.93 & 0.95 & 0.93 \\ 
  0.10 & 0.25 & 100 & 0.03 (0.11) & -3.37 (-3.13) & 0.02 (1.76) & 0.94 & 0.96 & 0.95 \\ 
  0.10 & 0.25 & 250 & 0.03 (0.09) & -5.68 (-4.86) & 0.04 (1.71) & 0.94 & 0.19 & 0.94 \\ 
  0.10 & 0.25 & 1000 & 0.03 (0.09) & -6.87 (-4.89) & -0.02 (0.36) & 0.95 & 0.30 & 0.95 \\ 
   \hline 
0.10 & 0.50 &  50 & 0.06 (0.49) & -0.36 (-0.76) & 0.00 (1.57) & 0.92 & 0.93 & 0.93 \\ 
  0.10 & 0.50 & 100 & 0.13 (0.44) & -0.41 (-0.98) & 0.05 (1.60) & 0.94 & 0.94 & 0.94 \\ 
  0.10 & 0.50 & 250 & 0.12 (0.39) & -1.87 (-1.74) & 0.05 (1.40) & 0.94 & 0.94 & 0.94 \\ 
  0.10 & 0.50 & 1000 & 0.16 (0.36) & -3.78 (-2.75) & -0.04 (0.48) & 0.95 & 0.61 & 0.95 \\ 
   \hline 
0.20 & 0.25 &  50 & 0.03 (0.25) & -10.27 (-8.81) & 0.07 (3.28) & 0.92 & 0.28 & 0.94 \\ 
  0.20 & 0.25 & 100 & 0.07 (0.22) & -12.98 (-10.56) & 0.10 (2.89) & 0.94 & 0.16 & 0.95 \\ 
  0.20 & 0.25 & 250 & 0.06 (0.18) & -13.75 (-9.64) & 0.05 (0.73) & 0.94 & 0.31 & 0.95 \\ 
  0.20 & 0.25 & 1000 & 0.03 (0.04) & -0.09 (-1.45) & 0.01 (-0.50) & 0.95 & 0.86 & 0.95 \\ 
   \hline 
0.20 & 0.50 &  50 & 0.23 (0.93) & -3.07 (-3.00) & 0.11 (2.82) & 0.92 & 0.93 & 0.93 \\ 
  0.20 & 0.50 & 100 & 0.34 (0.87) & -6.37 (-4.50) & 0.15 (2.46) & 0.94 & 0.94 & 0.95 \\ 
  0.20 & 0.50 & 250 & 0.33 (0.72) & -7.66 (-5.52) & -0.01 (0.91) & 0.94 & 0.61 & 0.95 \\ 
  0.20 & 0.50 & 1000 & 0.10 (0.14) & -0.08 (-1.00) & 0.03 (-0.29) & 0.95 & 0.86 & 0.95 \\ 
   \hline 
0.50 & 0.25 &  50 & 0.18 (0.43) & -35.04 (-21.86) & -0.59 (-0.12) & 0.92 & 0.36 & 0.93 \\ 
  0.50 & 0.25 & 100 & 0.07 (0.21) & -1.75 (-10.78) & -0.10 (-1.70) & 0.94 & 0.68 & 0.94 \\ 
  0.50 & 0.25 & 250 & -0.03 (-0.01) & -0.05 (-0.54) & -0.02 (-0.27) & 0.94 & 0.93 & 0.94 \\ 
  0.50 & 0.25 & 1000 & 0.02 (0.01) & 0.01 (0.01) & 0.01 (0.01) & 0.94 & 0.94 & 0.94 \\ 
   \hline 
0.50 & 0.50 &  50 & 0.89 (1.68) & -20.33 (-12.93) & -0.74 (1.00) & 0.92 & 0.55 & 0.93 \\ 
  0.50 & 0.50 & 100 & 0.35 (0.78) & -2.33 (-7.00) & -0.10 (-0.62) & 0.94 & 0.68 & 0.94 \\ 
  0.50 & 0.50 & 250 & -0.04 (0.01) & -0.07 (-0.42) & -0.04 (-0.20) & 0.94 & 0.93 & 0.94 \\ 
  0.50 & 0.50 & 1000 & 0.03 (0.03) & 0.03 (0.03) & 0.03 (0.03) & 0.94 & 0.94 & 0.94 \\ 
   \hline 
 \hline
\end{tabular}
\end{table}

% \newpage 
% \bibliographystyle{apalike}
% \bibliography{refs}
% \printbibliography

\end{document}